\documentclass[10pt,leqno]{amsart}
\usepackage{graphicx}
\usepackage{indentfirst,csquotes}

\usepackage{amssymb,amsthm,amsmath}
\usepackage{xcolor,paralist,hyperref,titlesec,fancyhdr,etoolbox}

\usepackage{amssymb}
\usepackage{amsmath}
\usepackage{graphicx} 
\usepackage{eurosym}
\usepackage{amsfonts}
\usepackage{amsmath}
\usepackage{amsfonts}
\usepackage{epsfig}
\usepackage{graphicx}
\usepackage{afterpage}
\usepackage{fancyhdr}
\usepackage{subcaption}
\usepackage{hyperref}
\usepackage{mathrsfs}
\usepackage{bigints}

\usepackage{lipsum}

\def\vv{\boldsymbol{v}}
\def\ev{\boldsymbol{e}}
\def\Dv{\boldsymbol{D}}
\def\Av{\boldsymbol{A}}
\def\Sv{\boldsymbol{S}}
\def\Tv{\boldsymbol{T}}
\def\Iv{\boldsymbol{I}}
\def\d{\mathrm{d}}
\def\eps{\epsilon}
\def\F{\mathscr{F}}

\def\heta{\widehat{\eta}}
\def\veps{\varepsilon}

\def\be{\begin{equation}}
\def\en{\end{equation}}
\def\iid{|II_{\Dv}|}
\def\iis{|II_{\Sv}|}
\def\iia{II_{\Av}}

\usepackage{titlesec}
\titleformat{\section}[block]{\normalfont\bfseries}{\thesection}{0.6em}{}
\titlespacing*{\section}{0em}{*2}{*1.2}
\titleformat{\subsection}[block]{\normalfont\bfseries\itshape}{\thesubsection}{0.6em}{}
\titlespacing*{\subsection}{4em}{*1}{*0.6}

\hypersetup{ colorlinks=true, linkcolor=black, filecolor=black, urlcolor=black }

\usepackage{lipsum}

\begin{document}
\title{Viscoplastic flows in narrow channels: Herschel-Bulkley models versus its regularizations} 
\author[B. Calusi]{ Benedetta Calusi}
\address{Dipartimento di Matematica e Informatica “Ulisse Dini”, Universit\`a degli Studi di Firenze, Viale Morgagni
67/a, 50134, Firenze, Italy.}
\email{benedetta.calusi@unifi.it}
\author[A. Farina]{Angiolo Farina}
\address{Dipartimento di Matematica e Informatica “Ulisse Dini”, Universit\`a degli Studi di Firenze, Viale Morgagni
67/a, 50134, Firenze, Italy.}
\email{angiolo.farina@unifi.it}
\author[L. Fusi]{Lorenzo Fusi}
\address{Dipartimento di Matematica e Informatica “Ulisse Dini”, Universit\`a degli Studi di Firenze, Viale Morgagni
67/a, 50134, Firenze, Italy.}
\email{lorenzo.fusi@unifi.it}
\author{ L. Vergori}
\address{Dipartimento di Ingegneria, Universit\`a degli Studi di Perugia, via Goffredo Duranti 93, 06125, Perugia, Italy}
\email{luigi.vergori@unipg.it}

\let\thefootnote\relax

\begin{abstract}
We investigated the two-dimensional flows of a viscoplastic fluid in symmetric channels with impermeable walls under no-slip boundary conditions. As response functions for the Cauchy stress tensor of the viscoplastic fluid, we considered both the celebrated Herschel-Bulkley model and a very general class $\mathcal{C}$ of its regularizations that depend on a positive parameter with the same physical dimensions as the strain rate and known as the \emph{regularization parameter}.  Within this class of regularized Herschel-Bulkley models,  the response function for the viscosity of the viscoplastic fluid tends to the non-smooth Herschel-Bulkley viscosity function as the regularization parameter tends to zero.  To make the equations governing the flow amenable to analysis, we considered channels with small aspect ratio so that the lubrication approximation can be used. In this way, we were able to obtain analytical solutions, perform an asymptotic analysis of the regularized solutions, and compare the results predicted by the Herschel-Bulkley model and its regularizations. We found that for any given channel, the regularized flows predicted by the regularizations in $\mathcal{C}$ tend to the same velocity field in the limit as the regularization parameter tends to zero. Such an asymptotic flow coincides with that predicted by the Herschel-Bulkley model only if the viscoplastic fluid flows in plane channels. Instead, in channels with curved walls, the results are markedly different.
\end{abstract} 

\maketitle

\bigskip


\section{Introduction}
Mineral slurries, paints, suspensions, crude oils, extrusions in 3D printing and many other materials commonly used in industrial applications exhibit a critical threshold for the stress (usually called the \emph{yield stress}) below which they behave like a rigid body, while they flow like a fluid if the shear stress exceeds the yield stress. These materials are often called viscoplastic or yield-stress fluids \cite{Bird1983}.  

Quite recently, there has been renewed increasing interest in the rheology of non-Newtonian fluids, as attested by numerous theoretical (e.g., \cite{BALMFORTH2004,Falsaperla2020,Farina2018,FernndezNieto2010,Fusi2023,Gianni2022,Rajagopal2012}) and experimental (e.g., \cite{Allouche2017,Coussot2014,Dinkgreve2018,Dinkgreve2016,FORTERRE2003,MounkailaNoma2021}) studies carried out over the past few decades. In particular, various mathematical models for the Cauchy stress tensor of viscoplastic fluids have been proposed. The most widely used and well-established ones are those of Bingham \cite{Bingham1922}, Casson \cite{Casson1959}, and Herschel \& Bulkley \cite{Herschel1926}.


Despite their widespread use, the existence of a real yield stress remains a subject of ongoing rheological debate \cite{Astarita1990,Barnes1985,Barnes1999,Frigaard2005,Frigaard2017}. Although yield stress models are idealized representations of real materials, they often introduce analytical and numerical difficulties that can lead to physically ambiguous results and even paradoxes \cite{Fusi2014,Fusi2015}. A well-known example is the so-called lubrication paradox concerning the existence or non-existence of real unyielded plug regions in yield-stress fluids that flow through non-plane channels with small aspect ratios. Lipscomb and Denn \cite{Lipscomb} were the first to arise interest in this question by arguing that real rigid plug
regions should not exist in non-plane channels. Specifically, in the framework of the lubrication approximation  in narrow channels with non-flat boundaries,  Lipscomb and Denn \cite{Lipscomb} determined a plug speed that varies slowly in the main flow direction. These regions  are called \emph{pseudo-plugs} and their boundaries are generally referred to as \emph{pseudo-yield surfaces}.  Putz \emph{et al.} \cite{Putz2009} were the first to realize that Lipscomb and Denn's deduction was incorrect, and real unyielded plugs can exist in complex geometries, too. 

Such issues do not arise when using regularized models. Thanks to the introduction of a positive regularization parameter (that is usually denoted $\epsilon$), regularized models are smooth approximations of the yield-stress behaviour and tend to the `exact' yield-stress model in the limit as $\epsilon \rightarrow 0^+$ \cite{Fusi2014,Fusi2015,Farina2024}. 

Recent studies have highlighted significant discrepancies between the predictions of `exact' yield-stress models and their regularized counterparts, particularly under lubrication approximations (e.g., \cite{Calusi2023,Calusi2022,Falsaperla2020,Farina2024,Fusi2022}). For instance, Farina \emph{et al.} \cite{Farina2024} showed that in channels with curved walls the flows predicted by regularized models differ markedly from those derived by using the `exact' Bingham model, even in the limit as $\epsilon \rightarrow 0^+$. In contrast, in channels with flat walls, at small enough values of $\epsilon$ the flows and yield surfaces predicted by regularized models match closely the `exact' ones. In addition, in the limit as the regularization parameter tends to zero the regularized and `exact' flows coincide. The asymptotic analysis in \cite{Farina2024} further reveals that, regardless of the specific regularization used, the asymptotic asymptotic yield surfaces and velocity fields depend solely on the geometry of the channel walls and the value of the Bingham number. Additionally, the differences between the models persist even when the full governing equations are solved numerically, clearly indicating that the lubrication approximation does not significantly affect these findings.

To our knowledge, no detailed analysis and comparisons of the results predicted by the Herschel–Bulkley model and its regularizations similar to those carried out in \cite{Farina2024} have been conducted yet. This lack represents the main motivation for our study.

The Herschel–Bulkley model is widely used to describe the rheology of non-Newtonian fluids that exhibit both yield stress behaviour and shear-dependent viscosity. It generalizes the Bingham model with the introduction of a power-law relationship between shear stress and shear rate; in such a way it offers greater flexibility in capturing complex fluid behaviours. In fact, denoted $n$ the (positive) power-law index, the Herschel–Bulkley model captures the shear-thinning behaviour if $0<n<1$, or the shear-thickening behaviour if $n>1$. With $n=1$, the Herschel–Bulkley model reduces to the Bingham model. For recent studies on Herschel–Bulkley fluids we refer the readers to \cite{Burgos1999,FernndezNieto2023,FUSI2017,Panaseti2018,Panaseti2019}.

The aims of this paper are three-fold. First, being inspired by the methodology and analytical approach developed in \cite{Farina2024}, we study the two-dimensional flows of a viscoplastic fluid in narrow symmetric channels within the framework of the lubrication approximation. To do so, we use both the `exact' and regularized Herschel–Bulkley models. Second, we analyse the asymptotics of the regularized flows and pseudo-plugs in the limit as $\epsilon \rightarrow 0^+$. Third, we compare these asymptotic predictions with those obtained with the `exact' model. Our results reveal that, in channels with non-flat boundaries, the asymptotic flows differ significantly from the `exact' ones. Moreover, for any given profile of the walls that bound the channel the asymptotic flows do not depend on the specific regularization used. In particular, when $n=1$ we retrieve the results obtained by Farina \emph{et al.} \cite{Farina2024} for Bingham fluids.

The structure of the paper is as follows. In Section \ref{Sec:PbForm-HB}, we present the Herschel–Bulkley model and its regularized versions. In Section \ref{Steady-2D} we derive the equations that govern the steady two-dimensional flows of a viscoplastic fluid in a narrow symmetric channel within the framework of the lubrication approximation. Assuming that the boundaries of the channel are impermeable and no-slip, we determine the `exact' and regularized two-dimensional flows in Sections \ref{Approx-2D} and  \ref{Sec:PbForm-RHB}, respectively. We then discuss the behaviour of the regularized flows in the limit as the regularization parameter tends to zero  and compare the asymptotic results with those predicted by the  `exact' Herschel-Bulkley model. We finally conclude with some remarks (Section \ref{Sec:Dis+Conc}).

\section{ Herschel-Bulkley model and its regularizations}\label{Sec:PbForm-HB} 

The Cauchy stress tensor $\Tv$ of an incompressible fluid can be expressed as the sum of an arbitrary spherical stress term, $-p\Iv$,  with the Lagrange multiplier associated with the incompressibility constraint $p$ having the same physical dimensions as the pressure, and a traceless extra stress tensor $\Sv$ (usually called the deviatoric part of the Cauchy stress tensor) given by a constitutive relation. In formulae, the Cauchy stress tensor reads $\Tv=-p\Iv+\Sv$. 

The mechanical response of some  viscoplastic fluids is described by the  Herschel-Bulkley model according to which the extra stress is related to the strain rate tensor $\Dv=[\nabla\vv+(\nabla\vv)^T]/2$, with $\vv$ denoting the velocity field, through
\begin{equation}\label{hbmod}
\left\{\begin{array}{ll}
\Sv=2\underbrace{\left(2^{n-1}K\iid^{(n-1)/2}+\dfrac{\tau_0}{2\iid^{1/2}}\right)}_{\displaystyle=\eta_{HB}(\iid^{1/2})}\Dv , \quad & \textrm{if } \iis^{1/2}>\tau_0,\\
[4mm]
\Dv=\boldsymbol{0} , \quad & \textrm{if } \iis^{1/2}\leq \tau_0,
\end{array}\right.
\end{equation}
where the positive parameters $K$ and $n$ are, respectively,  the consistency and (dimensionless) power-law indices, $\tau_0>0$ denotes the yield stress  and, for any traceless tensor $\Av$, $\iia=-|\Av|^2/2$ is the second principal invariant of $\Av$.

The constitutive model \eqref{hbmod} tells us that the deviatoric stress tensor $\Sv$ can be related to the rate of strain tensor $\Dv$ proving that $\iis$ exceeds a critical threshold: the yield stress $\tau_0$. Below this critical threshold, the motion of the viscoplastic  fluid is rigid. As a direct consequence of this, the effective viscosity $\eta_{HB}$ is well defined only in the yielded region where $\iis>\tau_0$. In the unyielded region where $\iis\leq \tau_0$ the effective viscosity is infinite. Moreover, at large enough strain rates (more precisely for $|\Dv|=2\iid^{1/2}>(\tau_0/K)^{1/n}$) the mechanical behaviour of a Herschel-Bulkley fluid resembles that of a shear-thinning fluid if $0<n<1$, a Newtonian fluid if $n=1$ and a shear-thickening fluid if  $n>1$.  Note that if $n=1$ in \eqref{hbmod}, then the Herschel-Bulkley response function for the extra stress  reduces to the Bingham model \cite{Bingham1922}.

The non-smooth nature of the effective viscosity  $\eta_{HB}$ prevents straightforward numerical computations \cite{Frigaard2005} and leads to poor curve fitting when modelling data from viscoplastic fluids \cite{SouzaMendesDutra}. One possible way to address these issues is to introduce regularized response functions for the viscosity (and consequently for the deviatoric stress tensor $\Sv$).

By a regularized model it is usually meant a constitutive relation for the deviatoric stress in terms of the rate of strain tensor of the form 
\begin{equation}\label{hbgen}
\Sv=2\eta(\iid^{1/2};\eps)\Dv,
\end{equation}
where the positive constant  $\eps$ has  the same physical dimensions as the rate of strain is usually referred to as the \emph{regularization parameter}.   The effective viscosity   $\eta(\cdot;\eps)$   is  a smooth  function defined over the interval $[0,+\infty[$ such that for any fixed $\eps>0$
\begin{subequations}\label{regcond}
\be\label{limreg}
\lim_{\xi\rightarrow0^+}\eta(\xi;\eps)=\dfrac{\tau_0}{2\eps},\qquad \lim_{\xi\rightarrow+\infty}\,\dfrac{\eta(\xi;\eps)}{2^{n-1}K\xi^{n-1}}=1,
\en
and
\be\label{increta}
\dfrac{\d }{\d \xi}\Big[\xi\eta(|\xi|;\eps)\Big]>0 ,
\en
for all $\xi\in\mathbb{R}$, while in the limit as $\eps\rightarrow0^+$, $\eta(\cdot;\eps)$ tends pointwise to the viscosity function $\eta_{HB}$   in $[0,+\infty[$, namely
\be\label{reqagg}
\lim_{\eps\rightarrow0^+}\eta(\xi;\eps)=\left\{\begin{array}{ll}
\eta_{HB}(\xi) , \quad &\textrm{if } \xi>0,\\
[5mm]
+\infty , \quad &\textrm{if } \xi=0,
\end{array}\right.
\en
and uniformly to $\eta_{HB}$  in $[a,+\infty[$ for all $a>0$, \emph{viz}
\be\label{reqaggun}
\lim_{\eps\rightarrow0^+}\sup_{\xi\in[a,+\infty[}\big|\eta(\xi;\eps)-\eta_{HB}(\xi)\big|=0 \quad \textrm{for all } a>0.
\en
\end{subequations}

Unlike the Herschel-Bulkley model  \eqref{hbmod}, the regularized constitutive relations \eqref{hbgen} are not based on the notion of yield stress. The regularization parameter $\eps$ is taken small (in the sense that $0<\eps(K/\tau_0)^{1/n}\ll1$) so that, in view of \eqref{limreg}, the effective viscosity is very large but \emph{finite}  at infinitesimally small strain rates. Only in the limit as $\eps\rightarrow0^+$ the effective viscosity diverges as the strain rate approaches zero. In this limit, the regularized constitutive relation \eqref{hbgen} converges pointwise to the Herschel–Bulkley model \eqref{hbmod}  (cf. \eqref{reqagg}). The second limiting condition in \eqref{limreg} ensures that, as in the Herschel–Bulkley model, at sufficiently high strain rates the mechanical behaviour of a viscoplastic fluid described by the regularized model \eqref{hbgen} resembles that of a shear-thinning, Newtonian, or shear-thickening fluid, depending on whether  $0 < n < 1$, $n = 1$, or $n > 1$, respectively. Finally, condition \eqref{increta} implies that in shear flows, the shear stress is a monotonically increasing function of the shear rate.

In the literature, there are several regularizations of the Herschel-Bulkley model in the form \eqref{hbgen}, with the regularized viscosity function satisfying all the requirements \eqref{hbgen}. These include
\begin{itemize}
    \item the modified Herschel-Bulkley-Papanastasiou model, 
    \begin{equation}\label{papa}
 \Sv=2\underbrace{\left[\dfrac{1-\exp(-\iid^{1/2}/\eps)}{\iid^{1/2}}\left(2^{n-1}K\iid^{n/2}+\dfrac{\tau_0}{2}\right)\right]}_{\displaystyle=\eta_{SMD}(\iid^{1/2};\eps)}\Dv,
    \end{equation}
proposed by Souza Mendes and Dutra \cite{SouzaMendesDutra}   to prevent the blow up of the apparent viscosity as $\iid^{1/2}\rightarrow0$ also for Herschel-Bulkley fluids  with power-law index $n<1$  \cite{Mitsoulis,DosSantos,lovato};
\item the generalization of the Bercovier-Engelman regularization for viscoplastic  fluids introduced in \cite{FernndezNieto2023}
\begin{equation}\label{beren}
 \Sv=2\underbrace{\left(\dfrac{2^nK\iid^{n/2}+\tau_0}{2\sqrt{\iid+\eps^2}}\right)}_{\displaystyle=\eta_{BE}(\iid^{1/2};\eps)}\Dv;
\end{equation}
\item more generally, any  model in the form
\begin{equation}\label{genmod}
\Sv=2f(\iid^{1/2};\eps)\left(2^{n-1}K\iid^{n/2}+\dfrac{\tau_0}{2}\right)\Dv,
\end{equation}
where $f(\cdot;\eps)$ is a positive smooth function defined over the interval $[0,+\infty[$ such that for any fixed $\eps>0$
\begin{subequations}\label{reqf}
\begin{equation}
    \lim_{\xi\rightarrow0^+}f(\xi;\eps)=\dfrac1\eps, \quad \lim_{\xi\rightarrow+\infty}f(\xi;\eps)=0,
\end{equation}
\be
\dfrac{\d}{\d \xi}\Big[\xi f(|\xi|;\eps)\Big]>0 ,
\en
for all $\xi\in\mathbb{R}$, and in the limit as $\eps\rightarrow0^+$ the regularizing function $f$ satisfies the limiting conditions
\be
\lim_{\eps\rightarrow0^+}f(\xi;\eps)=\left\{\begin{array}{ll}
\dfrac1\xi , \quad &\textrm{if } \xi>0,\\
[5mm]
+\infty , \quad &\textrm{if } \xi=0,
\end{array}\right.
\en
and for all $a>0$,
\be
\lim_{\eps\rightarrow0^+}\sup_{\xi\in[a,+\infty[}\left|f(\xi;\eps)-\dfrac1\xi\right|=0.
\en
\end{subequations}
\end{itemize}

Both the modified Herschel-Bulkley-Papanastasiou model \eqref{papa} and the generalized Bercovier-Engelman regularization \eqref{beren} belong to the class of regularized models \eqref{genmod}. On the other hand, also the regularizing function in  the generalized Bercovier-Engelman model \cite{Fusi2022},
\begin{equation}\label{genberen}
f(\xi;\eps)=\dfrac{1}{(|\xi|^m+\eps^m)^{1/m}} \quad (m>0),
\end{equation}
satisfies all the requirements \eqref{reqf}, whence constitutive relations for the deviatoric stress of the form \eqref{genmod}, with $f(\cdot;\eps)$ as in \eqref{genberen},  are regularizations of the Herschel-Bulkley model \eqref{hbmod}.

It can be shown that the smoothness of the function $f(\cdot;\eps)$ can be weakened with the assumption of piecewise smoothness. In the following sections, for simplicity, we shall consider only smooth regularizations of the Herschel-Bulkley model. However, for illustration, the constitutive model \eqref{genmod}, with  
\begin{equation}\label{step}
f(\xi;\eps)=\dfrac{1}{\eps\max\left\{1,\dfrac{|\xi|}{\eps}\right\}}  ,
\end{equation}
as in the local regularization proposed by Gonz\'alez-Andrade and M\'endez Silva \cite{GONZALEZANDRADE},
is a regularization of the Hershel-Bulkley model that satisfies all the requirements \eqref{regcond}.

\section{Steady two-dimensional flows in a symmetric narrow channel\label{Steady-2D}}

\begin{figure}
    \centering
    \includegraphics[width=1\linewidth]{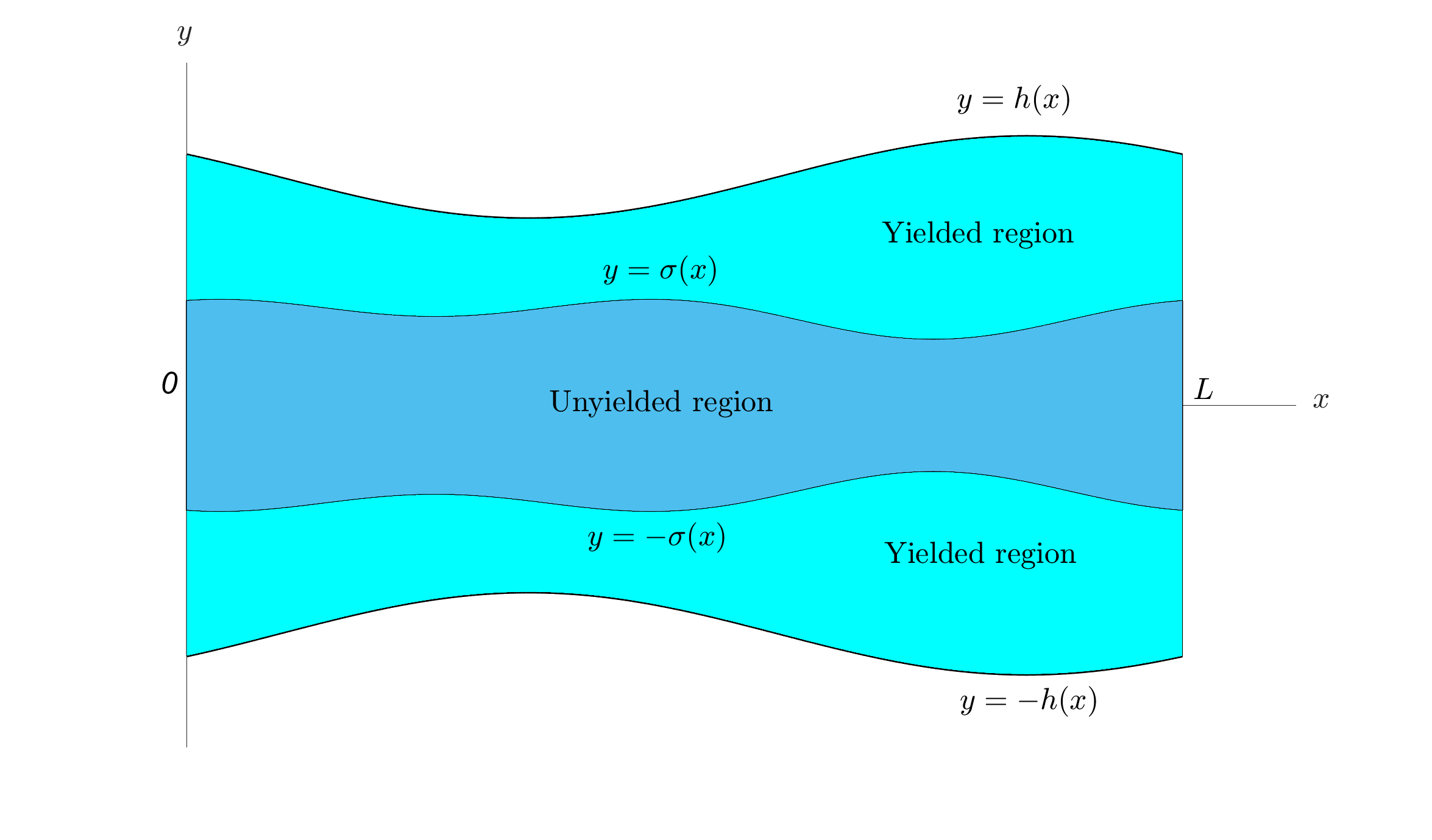}
    \caption{A schematic of a yield-stress fluid flow in a symmetric channel with curved walls of length $L$. The interface $y=\pm \sigma (x)$ separates the yielded region ($\iis^{1/2} > \tau_0$) and the unyielded one ($\iid^{1/2} \leq \tau_0$).  }
    \label{fig:schematic}
\end{figure}

In the sequel, we shall study the steady two-dimensional flow of a viscoplastic fluid in a symmetric channel (see Figure \ref{fig:schematic})
\begin{equation}\label{omega}
\Omega=\{(x,y)\in\mathbb{R}^2:0\leq x\leq L, \, 0\leq|y|\leq h(x)\},
\end{equation}
where $h(x)$ is a smooth positive function defined over the interval $[0,L]$.

In the absence of body forces, assuming that the velocity field is in the form $\vv(x,y)=u(x,y)\ev_1+v(x,y)\ev_2$, and using the subscript notation for differentiation with respect to the spatial variables $x$ and $y$,  the equations that govern the steady two-dimensional flows of a viscoplastic fluid in $\Omega$ read
\begin{equation}\label{goveq}
\left\{\begin{array}{ll}
\rho (uu_x+vu_y)=-p_x+{S_{11}}_x+{S_{12}}_y,\\
[5mm]
\rho (uv_x+vv_y)=-p_y+{S_{12}}_x+{S_{22}}_y,\\
[5mm]
u_x+v_y=0,
\end{array}\right.
\end{equation}
where $\rho$ is the (constant) mass density and the components $S_{ij}$ $(i,j=1,2)$ of the deviatoric stress tensor are given by the response functions \eqref{hbmod} or \eqref{hbgen}.

We assume that the channel boundaries are impermeable and that the fluid does not slip there. To enforce such requirements, it suffices to impose the following boundary conditions,
\begin{equation}\label{bcond}
u(x,\pm h(x))=v(x,\pm h(x))=0 , \quad \textrm{ for all }x\in[0,L].
\end{equation}

In view of the homogeneous boundary conditions \eqref{bcond} and the symmetry of the channel, it is physically reasonable to expect that symmetric flows (for which the velocity field is such that $\vv(x,y)=\vv(x,-y)$ for all $y\in[0,h(x)]$ and $x\in[0,L]$) occur. This allows us to limit our analysis to the upper region of the channel $\Omega^+ = \{(x,y)\in\Omega:y\geq 0\}$ provided that we substitute the boundary conditions \eqref{bcond} with the following ones 
\begin{equation}\label{bc}
\left\{\begin{array}{ll}
u(x,h(x))=v(x,h(x))=0,\\
[5mm]
u_y(x,0)=0, \quad v(x,0)=0,
\end{array}\right.
\end{equation}
for all $x\in[0,L]$.

Taking into account the symmetry of the flow and  following similar arguments as in \cite{Farina2024},  one can prove that the half discharge  
\begin{equation}\label{discharge}
Q=\int_0^{h(x)}u(x,y)\d y ,
\end{equation} 
is constant.  We treat the discharge as a prescribed quantity and assume it to be strictly positive.

Next, denoted $H=\displaystyle\max_{x \in [0,L]}h(x)$ and  introduced the  aspect ratio of the channel $\delta = H/L$,  we focus our analysis to narrow channels for which  
\begin{equation}
    0<\delta  \ll 1.
    \label{Lubrication-approx}
\end{equation}

For the subsequent analysis it is convenient to non-dimensionalise the governing equations \eqref{goveq}, the boundary conditions \eqref{bc} and the response functions \eqref{hbmod} and \eqref{hbgen}. To this end, we take $U=Q/H$ as the characteristic speed and introduce the  following  dimensionless   quantities
\begin{equation}\label{adim}
\left.\begin{array}{cc}
\tilde{x}={\displaystyle{\frac{x}{L}}},\quad \tilde{y}={\displaystyle{\frac{y}{H}}}={\displaystyle{\frac{y}{\delta L}}},\quad
\tilde{u}={\displaystyle{\frac{u}{U}}},\quad \tilde{v}={\displaystyle{\frac{v}{\delta U}}},\\
[5mm]
\tilde{h}={\displaystyle{\frac{h}{H}}},\quad  \tilde{p}={\displaystyle{\frac{
\delta H^{n} }{ K  U^n }}}p, \quad \tilde{\eta}={\displaystyle\frac{H^{n-1}}{KU^{n-1}}}\eta,\quad  \tilde{\Sv}={\displaystyle{\frac{H^n
}{ K U^n  }}}\Sv,\quad \tilde{\Dv}={\displaystyle{\frac{H}{U}}}\Dv. 
\end{array}\right.
\end{equation}

Inserting these dimensionless quantities into \eqref{hbmod}, \eqref{hbgen} and \eqref{omega}--\eqref{discharge}, omitting the tildes for simplicity of notation and denoted
\be
Re=\dfrac{\rho H^n}{KU^{n-2}}, \quad Bm = \dfrac{\tau_0H^n}{KU^n}=\dfrac{\tau_0}{\rho U^2}Re, \quad \veps=\dfrac{2\eps H}{U}
\en
the Reynolds and Bingham numbers and the dimensionless regularization parameter, respectively, yield 
\begin{itemize}
\item the dimensionless parametrization of the channel
\be
\Omega=\{(x,y)\in\mathbb{R}^2:x\in[0,1],\, |y|\leq h(x)\},
\en
with $\displaystyle 0<h_{\min}=\min_{x\in[0,1]}h(x)\leq h(x)\leq \max_{x\in[0,1]}h(x)=1$ for all $x\in[0,1]$;
    \item the dimensionless governing equations, 
\begin{equation}\label{dimeq}
\left\{\begin{array}{ll}
\delta Re \left(uu_x+vu_y\right)=-p_x+\delta{S_{11}}_x+{S_{12}}_y,\\
[5mm]
\delta^3 Re\left(uv_x+vv_y\right)=-p_y+\delta^2{S_{12}}_x+\delta{S_{22}}_y,\\
[5mm]
u_x+v_y=0,
\end{array}\right.
\end{equation}
    for all $y\in[0,h(x)]$ and $x\in[0,1]$, where the component of the velocity field along $\ev_1$ is such that 
 \begin{equation}\label{disdim}
\int_0^{h(x)}u(x,y)\d y =1,
 \end{equation}
 and $u$ and $v$ satisfy the dimensionless boundary conditions
\begin{equation}\label{dimbc}
    \left\{\begin{array}{ll}
    u(x,h(x))=v(x,h(x))=0,\\
    [5mm]
    u_y(x,0)=v(x,0)=0,
    \end{array}\right.
\end{equation}
for all $x\in[0,1]$;
\item the dimensionless Herschel-Bulkley model
\begin{equation}\label{dimhb}
\left\{\begin{array}{ll}
\Sv=2\left(2^{n-1}\iid^{(n-1)/2}+\dfrac{Bm} {2\iid^{1/2}}\right)\Dv , \quad &\textrm{if } \iis^{1/2}>Bm,\\
[5mm]
\Dv=\boldsymbol{0} , \quad &\textrm{if } \iis^{1/2}\leq Bm;
\end{array}\right.
\end{equation}
\item the dimensionless regularization
\begin{equation}\label{dimreg}
\Sv=2\eta\left(\iid^{1/2};\dfrac\veps2\right)\Dv.
\end{equation}
\end{itemize}

In \eqref{dimhb} and \eqref{dimreg} the dimensionless rate of strain tensor reads
\be
\Dv=\delta u_x\ev_1\otimes\ev_1+\dfrac12(u_y+\delta^2 v_x)(\ev_1\otimes\ev_2+\ev_2\otimes\ev_1)+\delta v_y\ev_2\otimes\ev_2,
\en
by which the dimensionless invariant $\iid$ is found to be
\be\label{iid}
\iid=\dfrac12\left[\delta^2(u_x^2+v_y^2)+\dfrac12(u_y+\delta^2v_x)^2\right].
\en

Limiting our analysis to flows occurring at small Reynolds number, say $O(Re)\leq O(1)$, and in view of the smallness of the aspect ratio  \eqref{Lubrication-approx}, to the leading order, the dimensionless governing equations \eqref{dimeq}  reduce to 
\begin{equation}\label{adim-lub}
    \left\{\begin{array}{ll}
        {S_{12}}_y=p_x, \\ 
        [3mm]
        p_y=0, \\
        [3mm]
        u_x+v_y=0,
    \end{array}\right.
\end{equation}
where, when using a regularized model, the dimensionless shear stress/ shear strain relation becomes
\begin{equation}\label{su}
S_{12}=\eta\left(\dfrac{|u_y|}{2};\dfrac\veps2\right)u_y=\heta(|u_y|;\veps)u_y=\mathscr{F}(u_y;\veps).
\end{equation}

In contrast, when using the `exact' Herschel–Bulkley model, the shear stress can be expressed as a function of the shear strain rate only if its magnitude exceeds the Bingham number; otherwise, the motion is rigid. The arguments provided by Fusi \emph{et al.} \cite{Fusi2015}   to demonstrate that the motion of a Bingham fluid is rigid when the shear stress magnitude does not exceed the Bingham number can be easily adapted to Herschel–Bulkley fluids. Therefore, we have that
\begin{equation} \label{Def:const-law-adim-lub-2}
\left\{\begin{array}{ll}
        S_{12} =   |u_y|^{n-1}u_y +Bm\,\mathrm{sgn}(u_y) ,  \quad & \textrm{if } \,  |S_{12}(x,y)|> Bm ,  \\
        [5mm]
       u_x = u_y = v_x = v_y = 0, \quad & \textrm{if } \, |S_{12}(x,y)| \leq Bm.
    \end{array}\right.
\end{equation}

We conclude this section on the basic equations governing steady flows within the lubrication approximation by observing that the Lagrange multiplier depends only on $x$ (cf. \eqref{adim-lub}), and, since the discharge is positive, the fluid flows from the inlet to the outlet and, consequently, the component of the gradient of $p$ in the $x$-direction  $p_x$ is negative.

\section{Approximated two-dimensional flow in a Herschel-Bulkley fluid\label{Approx-2D}}

We now consider the symmetric steady two-dimensional flow of a Herschel-Bulkley fluid through a narrow channel by using the approximated equations derived in the previous section. In this case the shear stress can be expressed in terms of the strain rate only in the yielded region, whereas the stress is indeterminate in the unyielded plug region, i.e. in the two-dimensional domain
\be
\Omega_c=\{(x,y)\in\Omega:\,-\sigma(x)\leq y\leq\sigma(x)\},
\en
with $y= \pm \sigma(x)$ being the yield surfaces at which $|S_{12}(x,y)|$ equals the Bingham number. Due to symmetry in the sequel we limit our attention to the part of the unyielded region that lies in the half plane $y>0$,
\be
\Omega_c^+=\{(x,y)\in\Omega:\,0\leq y\leq\sigma(x)\}.
\en

On the other hand, in view of the smallness of the aspect ratio of the channel (cf. \eqref{Lubrication-approx}) and the symmetry of the flow, the following equation of balance of linear momentum must hold in the part of the fluid in the unyielded plug region (see \cite{Fusi2015} for more details) 
\be\label{bal}
\int_0^1[-\sigma(x)p_x(x,\sigma(x))+S_{12}(x,\sigma(x))]\d x=0.
\en

Equations \eqref{adim-lub} hold in the yielded region $\Omega^+\setminus\Omega_c$ where the shear stress exceeds the Bingham number in modulus. Then, from the second equation in \eqref{adim-lub}, the shear stress vs rate of shear strain relationship  valid in the yielded region \eqref{Def:const-law-adim-lub-2} and the negativeness of $p_x$  we deduce that also $u_y$ is negative. As direct consequences of this, the shear stress becomes
\be\label{s12y}
S_{12}=|u_y|^{n-1}u_y-Bm \quad \textrm{in } \,\Omega^+\setminus\Omega_c^+, 
\en
whence integrating the second equation in \eqref{adim-lub} yields 
\begin{equation}\label{uyhb}
     u_y = -\Big\{|p_x| [y-\sigma (x)]\Big\}^{1/n}.
\end{equation}
Next, integrating \eqref{uyhb} and taking into account the boundary conditions \eqref{dimbc}, the continuity of the velocity field at the yield surface and the constancy of $\vv$ in the plug unyielded region (as the motion is rigid there) we deduce that
\be\label{ucomp}
u(x,y)=\left\{\begin{array}{ll}
\dfrac{n|p_x (x)|^{\frac1n}}{n + 1} [h(x) - \sigma (x) ]^{1+\frac1n} \left\{ 1 - \left[\dfrac{y - \sigma (x)}{h(x) - \sigma (x)} \right]^{1+\frac1n} \right\} , \quad &\textrm{if }\, \sigma(x)<y\leq h(x),\\
[5mm]
\underbrace{\dfrac{n|p_x (x)|^{\frac 1n}}{n + 1} [h(x) - \sigma (x) ]^{1+\frac1n}}_{\displaystyle=u_c} , \quad &\textrm{if }\, 0\leq y\leq \sigma(x),
\end{array}\right.
\en
where $p_x$ and the yield surface $y=\sigma (x)$, though  yet unknown, are such that the plug-speed $u_c$ is constant. 

We now express $p_x$ and the yield surface in terms of the plug-speed $u_c$. To do this,  we use  \eqref{disdim} and \eqref{ucomp} to obtain
\begin{equation}
1=u_c\sigma(x)+\dfrac{n}{2n+1}|p_x(x)|^{\frac1n}\Big[h(x)-\sigma(x)\Big]^{ 2+\frac1n}=\dfrac{u_c}{2n+1}\Big[(n+1)h(x)+n\sigma(x)\Big],
\end{equation}
by which we deduce that
\begin{subequations}\label{sigpx}
\be\label{sighb}
\sigma(x)=\dfrac{2n+1}{nu_c}-\dfrac{n+1}{n}h(x) ,
\en
and
\be
|p_x(x)|=\dfrac{n(n+1)^n}{(2n+1)^{n+1}}\dfrac{u_c^{2n+1}}{[u_ch(x)-1]^{n+1}}.
\en
\end{subequations}

To complete the determination of the flow of a Herschel–Bulkley fluid in a narrow channel, it is sufficient to determine $u_c$. To this end, we first observe that, for the unyielded plug region to be simply connected and to extend from the inlet to the outlet, it is necessary and sufficient that $0<\sigma(x)<h(x)$ for all $x \in [0,1]$. This, in turn, is equivalent to requiring that the plug speed fulfils these inequalities
\be\label{ucint}
\dfrac{1}{h_{\min}}< u_c< \dfrac{2n+1}{n+1}, 
\en
which make sense provided that the profile of the upper boundary of the channel is such that
\be\label{hcondbm}
\dfrac{n+1}{2n+1}< h_{\min}\leq 1.
\en
From now on, we assume that condition \eqref{hcondbm} is met. Hence, from \eqref{bal}, \eqref{s12y} and \eqref{sigpx} we deduce that the plug-speed must satisfy the equation
\be\label{uceq}
\mathcal{H}(u_c)=\dfrac{(n+1)^nu_c^{2n}}{(2n+1)^{n+1}}\int_0^1\dfrac{2n+1-(n+1)u_ch(x)}{[u_ch(x)-1]^{n+1}}\d x=Bm.
\en

It can be easily proven that the function $\mathcal{H}$ is continuous and monotonically decreasing on the interval $\displaystyle\left]\dfrac{1}{h_{\min}},\dfrac{2n+1}{n+1}\right]$, and that
\be\label{hlim}
\lim_{\upsilon\rightarrow1/h_{\min}}\mathcal{H}(\upsilon)=+\infty.
\en
In view of the properties of $\mathcal{H}$, equation \eqref{uceq} admits a unique solution $u_c$ in the interval \eqref{ucint} if and only if
\be\label{bmcond}
Bm>\mathcal{H}\left(\dfrac{2n+1}{n+1}\right).
\en

\begin{figure}
\centering
\begin{subfigure}{0.45\textwidth}
    \includegraphics[width=\textwidth]{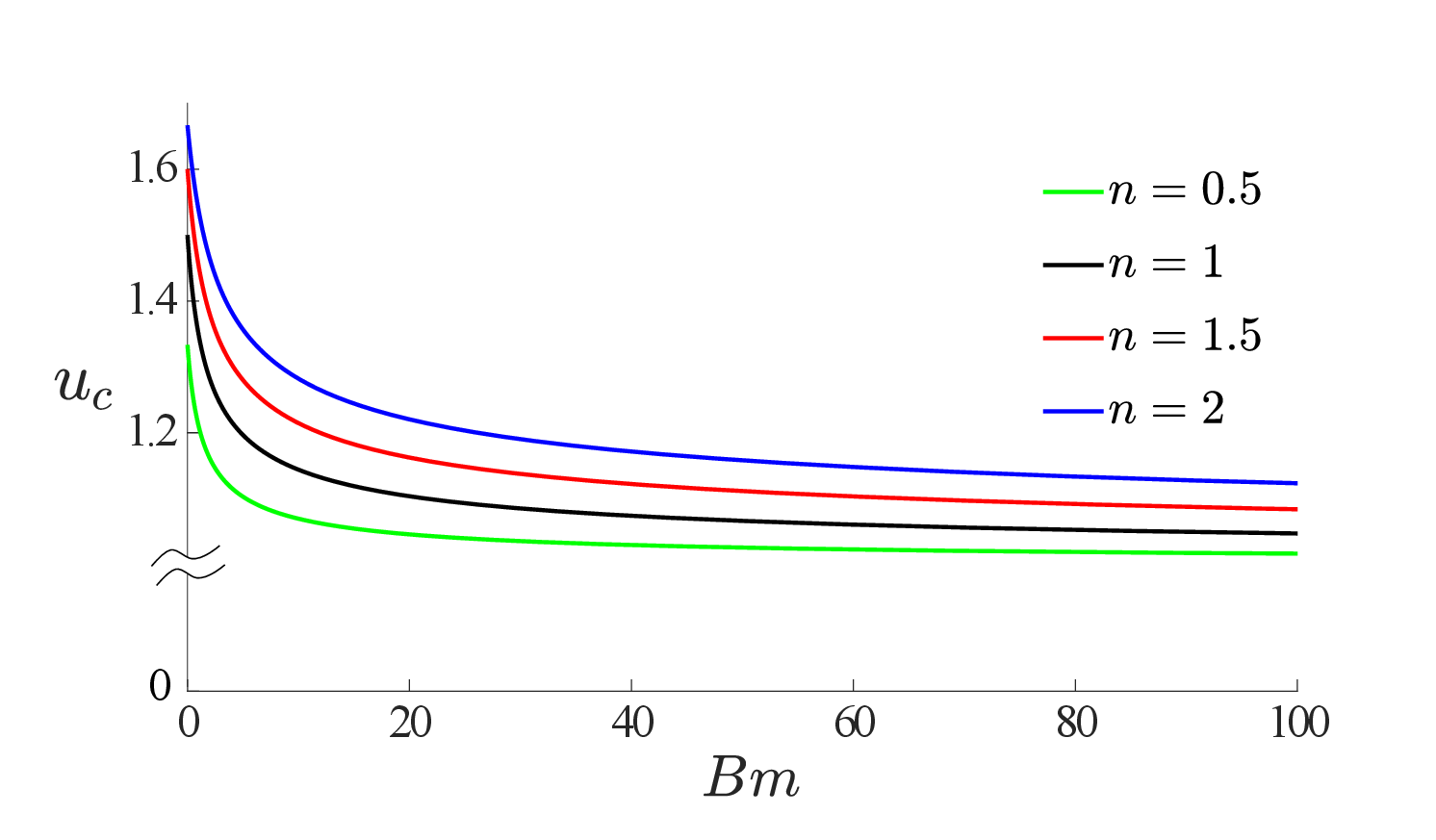}
    \caption{$h\equiv 1$.\label{sinpapa}}
\end{subfigure}
\hfill
\begin{subfigure}{0.45\textwidth}
    \includegraphics[width=\textwidth]{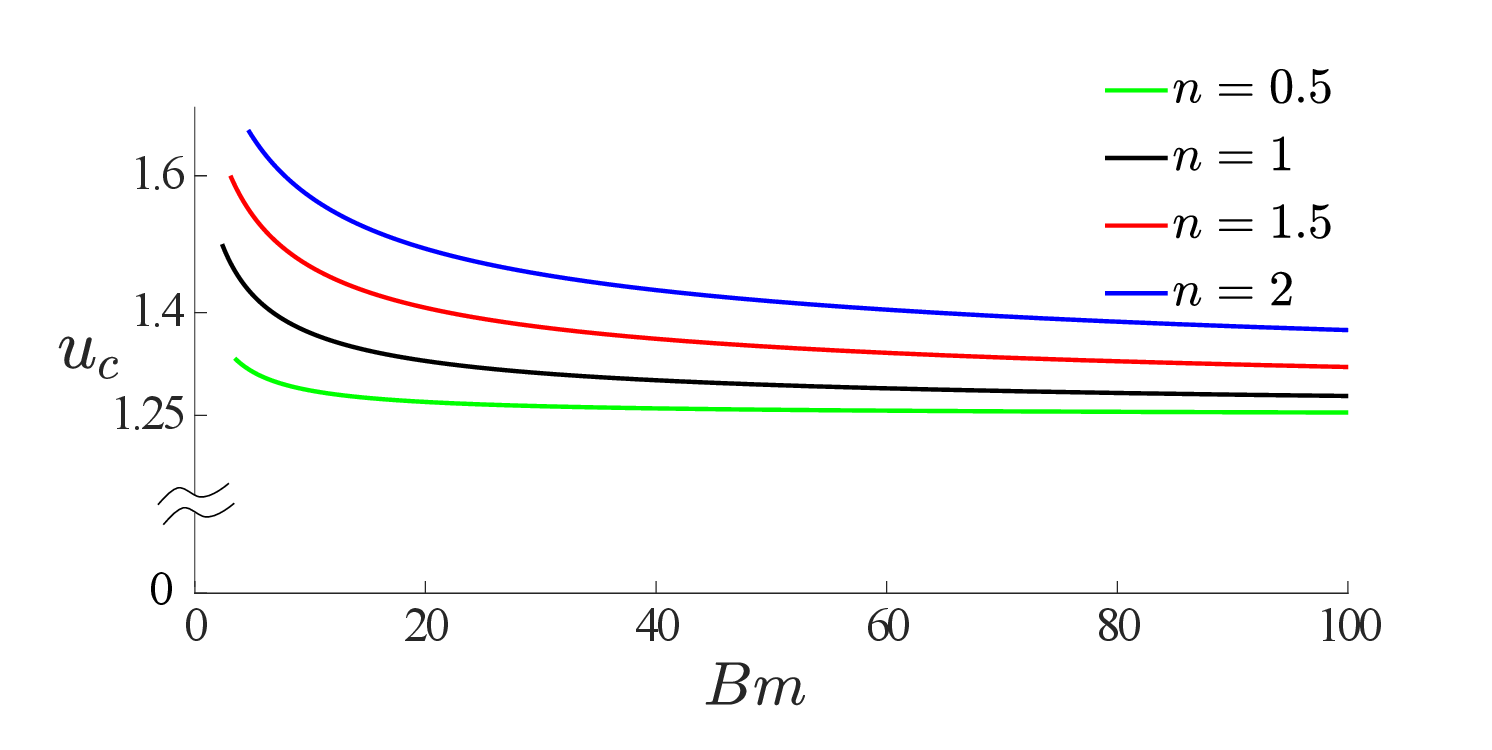}
    \caption{$h(x)=0.9+0.1\sin(2\pi x)$.\label{divpapa}}
   \end{subfigure}
    \begin{subfigure}{0.45\textwidth}
    \includegraphics[width=\textwidth]{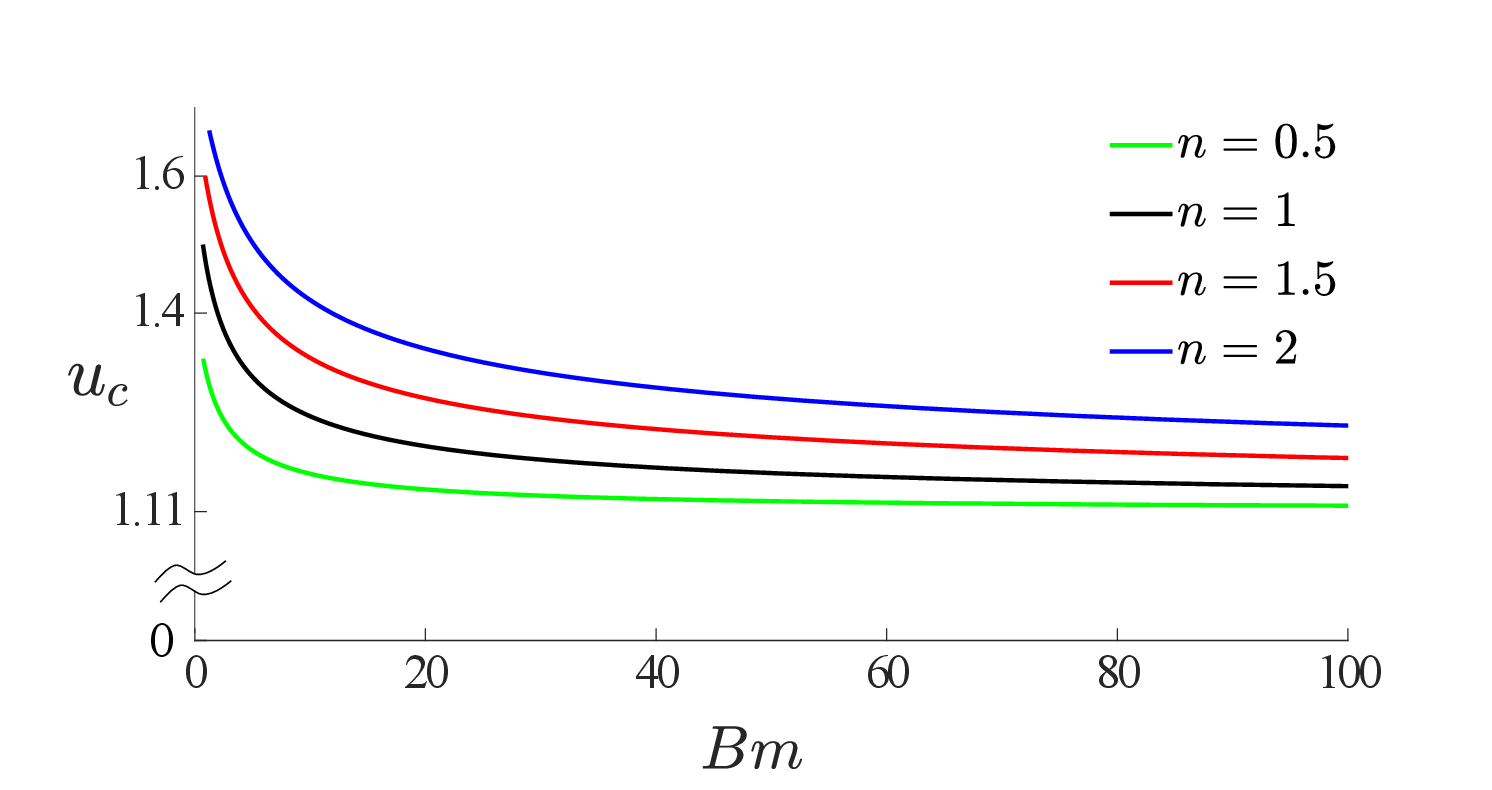}
    \caption{$h(x)=0.9+0.1(x-1)^2$.\label{conpapa}}
\end{subfigure}    
\hfill
 \begin{subfigure}{0.45\textwidth}
    \includegraphics[width=\textwidth]{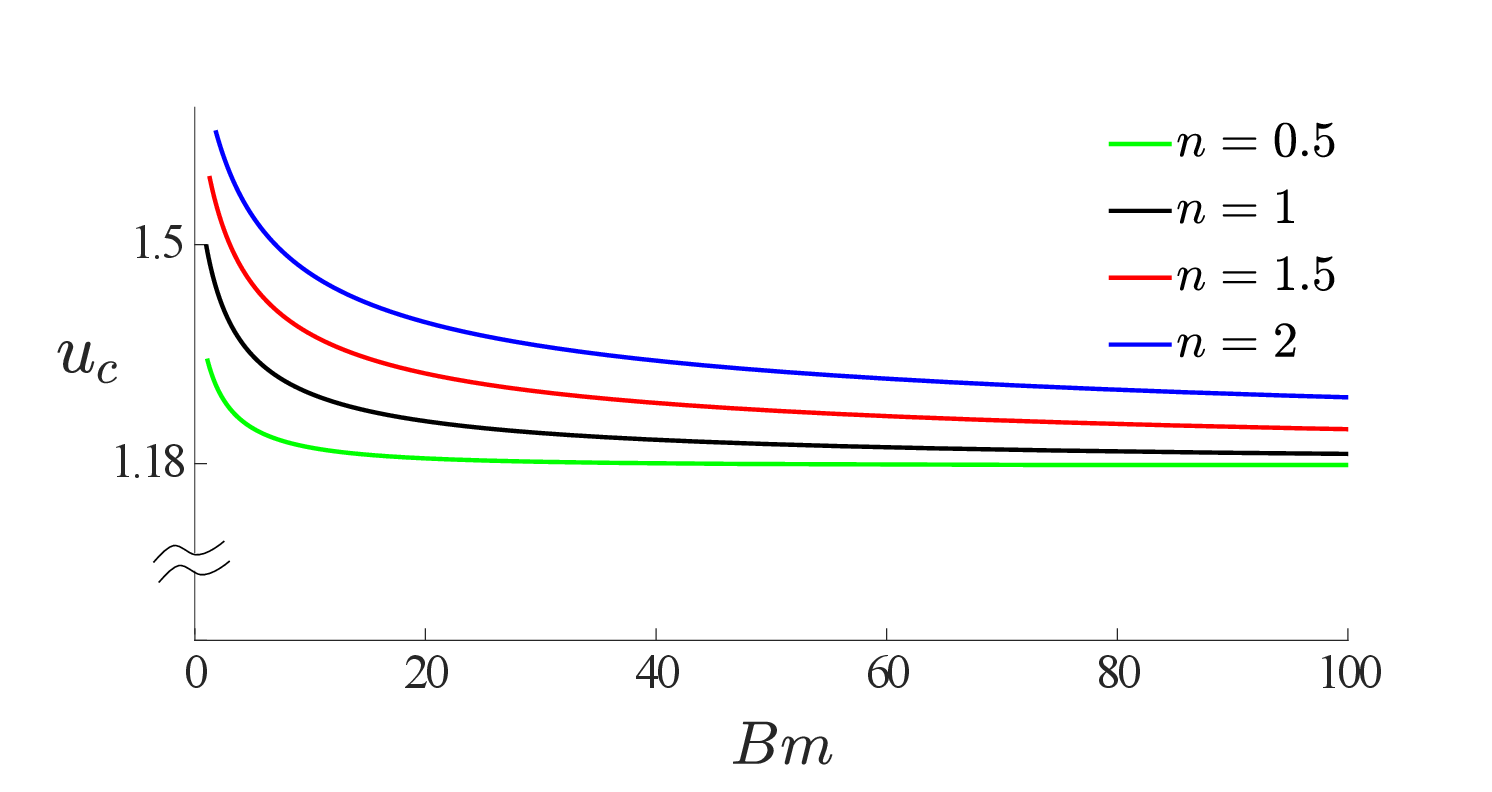}
    \caption{$h(x)=0.85+0.15x$.\label{uclin}}
\end{subfigure}  
\caption{Speeds of the plug region vs the Bingham number for some profiles of the upper boundary of the channel at different values of the power-law index $n$. For any given profile $y=h(x)$ of the upper boundary of   the channel,  the speed of the rigid core tends to $1/h_{\min}$ as $Bm\rightarrow +\infty$. \label{papat}}
\end{figure}

Figure \ref{papat} shows the plug speed $u_c$ as a function of the Bingham number for various profiles of channel walls. These plots clearly indicate that, regardless of the shape of the channel boundaries, the plug speed approaches $1/h_{min}$ as the Bingham number increases. This behaviour is a direct consequence of the invertibility of $\mathcal{H}$ and the limiting relation given in \eqref{hlim}.

In summary, we have shown that if the channel profile satisfies condition \eqref{hcondbm} and the Bingham number is sufficiently large (cf. \eqref{bmcond}), then the yield surface is simply connected and extends from the inlet to the outlet. The components of the velocity field are
\begin{subequations}\label{2de}
\be\label{uf}
u(x,y)=\left\{\begin{array}{ll}
u_c\left\{1-\left[\dfrac{y-\sigma(x)}{h(x)-\sigma(x)}\right]^{1+\frac1n}\right\} , \quad &\textrm{if }\, \sigma(x)<y\leq h(x),\\
[5mm]
u_c , \quad &\textrm{if }\, 0\leq y\leq \sigma(x),
\end{array}\right.
\en
\be\label{vf}
v(x,y)=\left\{\begin{array}{ll}
\left(1+\dfrac{1}{n}\right)u_c\,\dfrac{\Big[y-\sigma(x)\Big]^{1+\frac1n}\Big[h(x)-y\Big]}{\Big[h(x)-\sigma(x)\Big]^{2+\frac1n}}h_x(x) , \quad &\textrm{if }\, \sigma(x)<y\leq h(x),\\
[5mm]
0 , \quad &\textrm{if }\, 0\leq y\leq \sigma(x),
\end{array}\right.
\en
\end{subequations}
with $\sigma$ as in \eqref{sighb} and  $u_c=\mathcal{H}^{-1}(Bm)$. The component of the velocity field along $\ev_2$ has been determined from the third equation in \eqref{adim-lub}, the boundary conditions \eqref{dimbc} and \eqref{uf}.

\section{\label{Sec:PbForm-RHB} Approximated two-dimensional flow in a viscoplastic fluid with a  regularized Herschel-Bulkley viscosity function}

Within the lubrication approximation, we now determine the flow of a Herschel-Bulkley fluid through a symmetric narrow channel when the regularized constitutive relation for the shear stress is given by \eqref{su}.

As illustrations, when using the modified Herschel-Bulkley-Papanastasiou regularization \eqref{papa}, introduced by \cite{SouzaMendesDutra}, the approximated response function \eqref{su} reads
\be\label{papadim}
\mathscr{F}(u_y;\veps)=\heta_{SDM}(|u_y|;\veps)u_y=\dfrac{1-\exp(-|u_y|/\veps)}{|u_y|}(|u_y|^n+Bm)\,u_y,
\en
while for the generalisation of the Bercovier-Engelman regularization \eqref{beren} we have
\be
\mathscr{F}(u_y;\veps)=\heta_{BE}(|u_y|;\veps)u_y=\dfrac{|u_y|^n+Bm}{\sqrt{u_y^2+\veps^2}}\,u_y.
\en

In view of the properties of the regularized viscosity function \eqref{regcond}, for any $\veps>0$ $\mathscr{F}(\cdot;\veps)$  is an odd function defined over the whole real axis that vanishes at $\xi=0$ and fulfils the conditions at infinity  
\be
\lim_{\xi\rightarrow\pm\infty}\mathscr{F}(\xi;\veps)=\pm\infty.
\en
In addition, $\mathscr{F}(\cdot;\veps)$ is monotonically increasing for all $\veps>0$. Thus, it is invertible and its inverse $\mathscr{F}^{-1}(\cdot;\veps)$ satisfies the conditions
\be
\mathscr{F}^{-1}(0;\veps)=0, \quad \lim_{\zeta\rightarrow\pm\infty}\mathscr{F}^{-1},(\zeta;\veps)=\pm\infty ,
\en
 is continuously differentiable and its first derivative is positive as 
\be\label{derfinv}
\dfrac{\d}{\d \zeta}\mathscr{F}^{-1}(\zeta;\veps)=\dfrac{1}{|\mathscr{F}^{-1}(\zeta;\veps)|\,\heta'(|\mathscr{F}^{-1}(\zeta;\veps)|;\veps)+\heta(|\mathscr{F}^{-1}(\zeta;\veps)|;\veps)}>0 \quad \textrm{for all }\zeta\in\mathbb{R},
\en
where the prime denotes differentiation with respect to $|\xi|$.

Bearing in mind that $p_x$ is negative, from \eqref{dimbc} and \eqref{su} we deduce that, for any $x\in[0,1]$,   $u_y(x,y)$ is negative for all $0<y\leq h(x)$ and vanishes at $y=0$. This fact and the invertibility  of $\F(\cdot;\veps)$  allow us to integrate the first equation in \eqref{adim-lub} and thus determine the longitudinal velocity 
\be\label{ue}
u(x,y;\veps)=\int_y^{h(x)}\F^{-1}(|p_x(x)|\varsigma;\veps)\,\d \varsigma,
\en
where  $p_x$ is still as yet unknown. 

To determine  $p_x$,  we combine the dimensionless equation for the discharge \eqref{disdim} with \eqref{ue} and obtain 
\be\label{eqpg}
\int_0^{h(x)}\left[\int_y^{h(x)}\F^{-1}(|p_x(x)|\varsigma;\veps)\,\d\varsigma\right]\d y=1.
\en
Following similar arguments as in \cite{Farina2024},  one can prove that  there exists a unique continuously differentiable solution $\bar{p}_x(\cdot;\veps)$ whose derivative with respect to $x$ reads
\be\label{pxx}
\bar{p}_{xx}(x;\veps)=\dfrac{h(x)h_x(x)\F^{-1}(|\bar{p}_x(x;\veps)|h(x);\veps)}{\displaystyle \int_0^{h(x)}\left[\int_y^{h(x)}\dfrac{\varsigma}{\widehat{\eta}'(|\bar{p}_x(x;\veps)|\varsigma;\veps)|\bar{p}_x(x;\veps)|\varsigma+\widehat{\eta}(|\bar{p}_x(x;\veps)|\varsigma;\veps)}\d\varsigma\right]\d y}.
\en

From \eqref{derfinv}, \eqref{pxx} and the positiveness of $\F^{-1}(\cdot;\veps)$ in the interval $]0,+\infty[$ one deduces that $\mathrm{sgn}(\bar{p}_{xx}(x;\veps))=\mathrm{sgn}(h_{x}(x))$. In particular, $\bar{p}_{xx}(\cdot;\veps)$ vanishes identically if and only if $h\equiv1$. This means that if the channel is plane, then the component of the gradient of the Lagrange multiplier in the $x$-direction is constant everywhere.

To complete the determination of the two-dimensional flows, we insert $\bar{p}_x(\cdot;\veps)$ into \eqref{ue} and find the solution for the component $u$ of the velocity field, whence integrating the third equation in \eqref{adim-lub} yields the solution for the component of the velocity field along $\ev_2$:
\begin{align}\label{ve}
&v(x,y;\veps)=-\int_0^y u_x(x,y;\veps)\d y \\
\nonumber
&=h_x(x)\F^{-1}(\bar{p}_x(x;\veps)h(x);\veps)\\
\nonumber
& \times\left\{y-h(x)\dfrac{\displaystyle \int_0^{y}\left[\int_y^{h(x)}\dfrac{\varsigma}{\widehat{\eta}'(|\bar{p}_x(x;\veps)|\varsigma;\veps)|\bar{p}_x(x;\veps)|\varsigma+\widehat{\eta}(|\bar{p}_x(x;\veps)|\varsigma;\veps)}\d\varsigma\right]\d y}{\displaystyle \int_0^{h(x)}\left[\int_y^{h(x)}\dfrac{\varsigma}{\widehat{\eta}'(|\bar{p}_x(x;\veps)|\varsigma;\veps)|\bar{p}_x(x;\veps)|\varsigma+\widehat{\eta}(|\bar{p}_x(x;\veps)|\varsigma;\veps)}\d\varsigma\right]\d y}\right\}.
\end{align}

We now define the ``regularized'' pseudo-yield surface in $\Omega^+$, $y=\bar{\sigma}(x;\veps)$, as the locus at which $|S_{12}(x,\bar{\sigma}(x;\veps))|=Bm$. From \eqref{su} and \eqref{ue}, the regularized pseudo-yield surface is found to have equation
\be\label{se}
y=\bar{\sigma}(x;\veps)=\dfrac{Bm}{|\bar{p}_x(x;\veps)|}.
\en
Here, we use similar terminology as in \cite{Frigaard2004} and add the adjective ``pseudo'' to emphasize the fact that the motion of the part of the viscoplastic material that occupies the region $\Omega_\veps^+=\{(x,y)\in\Omega^+:0\leq y\leq \bar{\sigma}(x;\veps)\}$ is not rigid. To realize this, it suffices to differentiate \eqref{ue} with respect to $y$ to see that the shear rate $u_y$ does not vanish identically in $\Omega_\veps^+$.

If the Herschel-Bulkley fluid flows in a non-plane channel, the yield surface \eqref{se} differs significantly from the yield surface predicted by the `exact' Herschel-Bulkley model \eqref{sighb}. 
To prove this claim, it is sufficient to observe that $\mathrm{sign}({\sigma}_x)=-\mathrm{sign}(h_x)$ in $[0,1]$, while, in view of \eqref{pxx} and \eqref{se},  $\mathrm{sign}(\bar{\sigma}_x(x;\veps))=\mathrm{sign}(h_x(x))$ for all $x\in[0,1]$.  In simple terms, the `exact' yield surface has the opposite monotonicity of the profile of the upper wall of the channel, whereas the regularized pseudo-yield surface \eqref{se} has the same monotonicity as the boundary $y=h(x)$ (see Figures  \ref{papareg}, \ref{berreg}).  Instead, if the flow occurs in a plane channel (for which $h\equiv1$), the regularized pseudo-yield surface \eqref{se} is actually a yield plane (cf. \eqref{pxx}) like the `exact' yield plane \eqref{sighb}, with $h\equiv1$ (Figure \ref{papaberplane}). For any given regularization and any fixed value of the regularization parameter, the regularized pseudo-yield and `exact' yield planes do not coincide. However, in the next section we shall prove that, irrespective of the regularization used, the regularized pseudo-yield plane tends uniformly to the `exact' yield plane in the limit as $\veps\rightarrow0^+$.

\section{Asymptotic results for regularized Herschel-Bulkley models\label{Asympt-Results}}

The analysis presented in the previous section shows that, when adopting a regularization of the Herschel–Bulkley model in the form of \eqref{hbgen} for the stress–strain rate constitutive relation, and within the framework of the lubrication approximation, the equations governing two-dimensional flows in a symmetric narrow channel admit a unique solution for any value of the Bingham number. Given the pointwise (or, in some domains, uniform) convergence of the regularized viscosity in \eqref{hbgen} to the `exact' Herschel–Bulkley viscosity, it is natural to investigate the asymptotic behaviour of the two-dimensional flows, as well as the regularized pseudo-yield surfaces predicted by model \eqref{hbgen}, in the limit as $\veps$ tends to zero.

We start our asymptotic analysis by observing that in view of \eqref{reqagg}, 
\be\label{lim1}
\lim_{\veps\rightarrow0^+}\mathscr{F}(\xi;\veps)=\mathscr{F}_\star(\xi)=\left\{\begin{array}{ll}
\xi^n+Bm , \quad &\textrm{if } \,\xi>0,\\
[3mm]
0, \quad &\textrm{if } \,\xi=0,\\
[3mm]
-|\xi|^n-Bm , \quad &\textrm{if } \,\xi<0,
\end{array}\right.
\en
while from \eqref{reqaggun} and the oddness of $\mathscr{F}(\cdot;\veps)$ and $\mathscr{F}_\star$ we have that
\be\label{lim2}
\lim_{\veps\rightarrow0^+}\sup_{\xi\in[a,+\infty[}\Big|\F(\xi;\veps)-\xi^n-Bm\Big|=0 \quad \textrm{and} \quad \lim_{\veps\rightarrow0^+}\sup_{\xi\in]\infty, -a]}\Big|\F(\xi;\veps)+|\xi|^n+Bm\Big|=0  ,
\en
for all $a>0$. 

Then, from  \eqref{lim1} and \eqref{lim2}   we deduce the uniform convergence of  $\F^{-1}(\cdot;\veps)$ to
\be\label{gun}
g(\zeta)=\left\{\begin{array}{ll}
(\zeta-Bm)^{\frac1n}, \quad &\textrm{if }\, \zeta>Bm,\\
[3mm]
0, \quad &\textrm{if }\, |\zeta|\leq Bm,\\
[3mm]
-|\zeta+Bm|^{\frac1n}, \quad &\textrm{if }\, \zeta<-Bm
\end{array}\right.
\en
over the entire real axis. 

 Next, the change in  variable $p_x(x;\veps)\rightarrow-Bm/\sigma(x;\veps)$ in equation \eqref{eqpg} and the uniform convergence of $\F^{-1}(\cdot;\veps)$ to the function $g$ in \eqref{gun} imply the uniform convergence of $\bar{\sigma}(x;\veps)$ to the unique solution $\sigma_0(x)$ of equation 
\be\label{eqay}
\dfrac{n}{(2n+1)(n+1)}\left[\dfrac{Bm}{\vartheta(x)}\right]^{\frac1n}\Big[h(x)-\vartheta(x)\big]^{1+\frac1n}\left[(n+1)h(x)+n\vartheta(x)\right]=1 ,
\en
in the limit as $\veps\rightarrow0^+$. In other words, for any given regularization in the form \eqref{hbgen} the regularized pseudo-yield surface $y=\sigma(x;\veps)$ tends uniformly to the surface $y=\sigma_0(x)$.

We now exploit the uniform convergence of $\F^{-1}(\cdot;\veps)$ to $g$ and the uniform convergence of $\bar{p}_x(x;\veps)$ to $-Bm/\sigma_0(x)$ to deduce that, in the limit as $\veps\rightarrow0^+$, the velocity component $u(\cdot,\cdot;\veps)$ converges uniformly to the continuously differentiable function
\begin{subequations}\label{2d0}
\begin{align}\label{u0}
     u_0(x,y) &= \dfrac{n}{n+1}\left[\dfrac{Bm}{\sigma_0(x)} \right]^{\frac{1}{n}}[h(x)-\sigma_0 (x) ]^{1+\frac{1}{n}} \\
     \nonumber
     &\times 
     \left\{\begin{array}{ll}
         1,\quad  & \textrm{if } 0\leq y \leq \sigma_0 (x) , \\
         [3mm]
        1 -\left[\dfrac{y-\sigma_0 (x)}{h(x)-\sigma_0 (x)} \right]^{1+\frac{1}{n}} \quad & \textrm{if }  \sigma_0 (x)\leq y \leq h (x), 
     \end{array}\right.
\end{align}
and its first derivative $u_x(\cdot,\cdot;\veps)$  converges uniformly to ${u_0}_x$.
Finally, combining this result with \eqref{ve} yields that $v(\cdot,\cdot;\veps)$ tends uniformly to
\begin{align}\label{v0}
    v_0(x,y) 
    =& n\left[\dfrac{Bm}{\sigma_0(x)} \right]^{\frac{1}{n}}\dfrac{[h(x)-\sigma_0 (x) ]^{1+\frac{1}{n}}[2n\sigma_0(x)+h(x)]h_x(x)}{(n+1)h^2(x)+2nh(x)\sigma_0(x)+2n^2\sigma_0^2(x)}\\
    \nonumber
   &\times 
     \left\{\begin{array}{ll}
         y,\quad  & \textrm{if } 0\leq y \leq \sigma_0 (x) , \\
         [3mm]
        y -h(x)\left[\dfrac{y-\sigma_0 (x)}{h(x)-\sigma_0 (x)} \right]^{1+\frac{1}{n}} , \quad & \textrm{if }  \sigma_0 (x)\leq y \leq h (x).
     \end{array}\right.   
\end{align}
\end{subequations}

\begin{figure}[h!]
    \centering
    \begin{subfigure}{0.3\textwidth}
         \includegraphics[width=\textwidth]{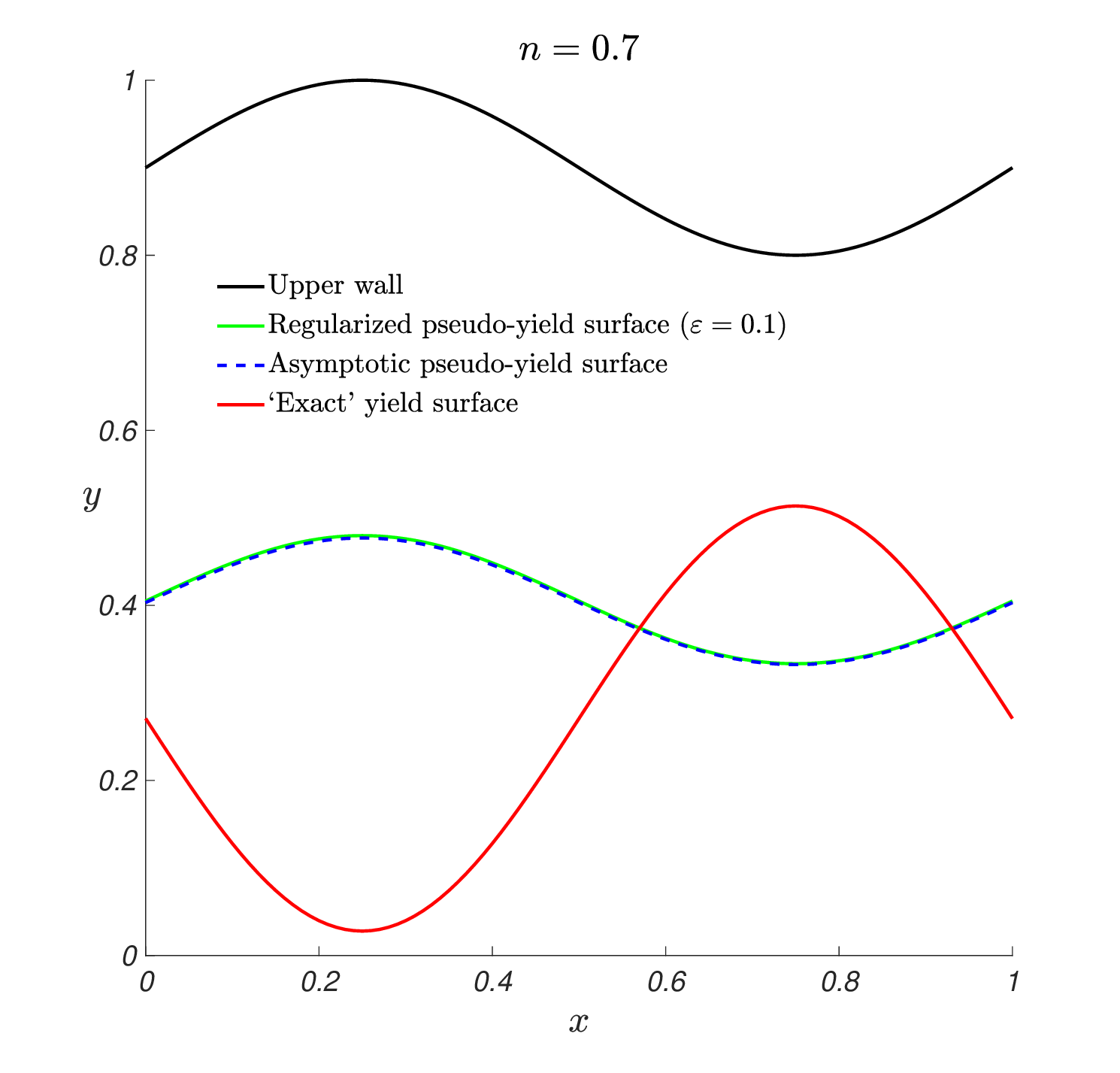}
         \caption{\label{pap07}}
         \end{subfigure}
         \begin{subfigure}{0.3\textwidth}
         \includegraphics[width=\textwidth]{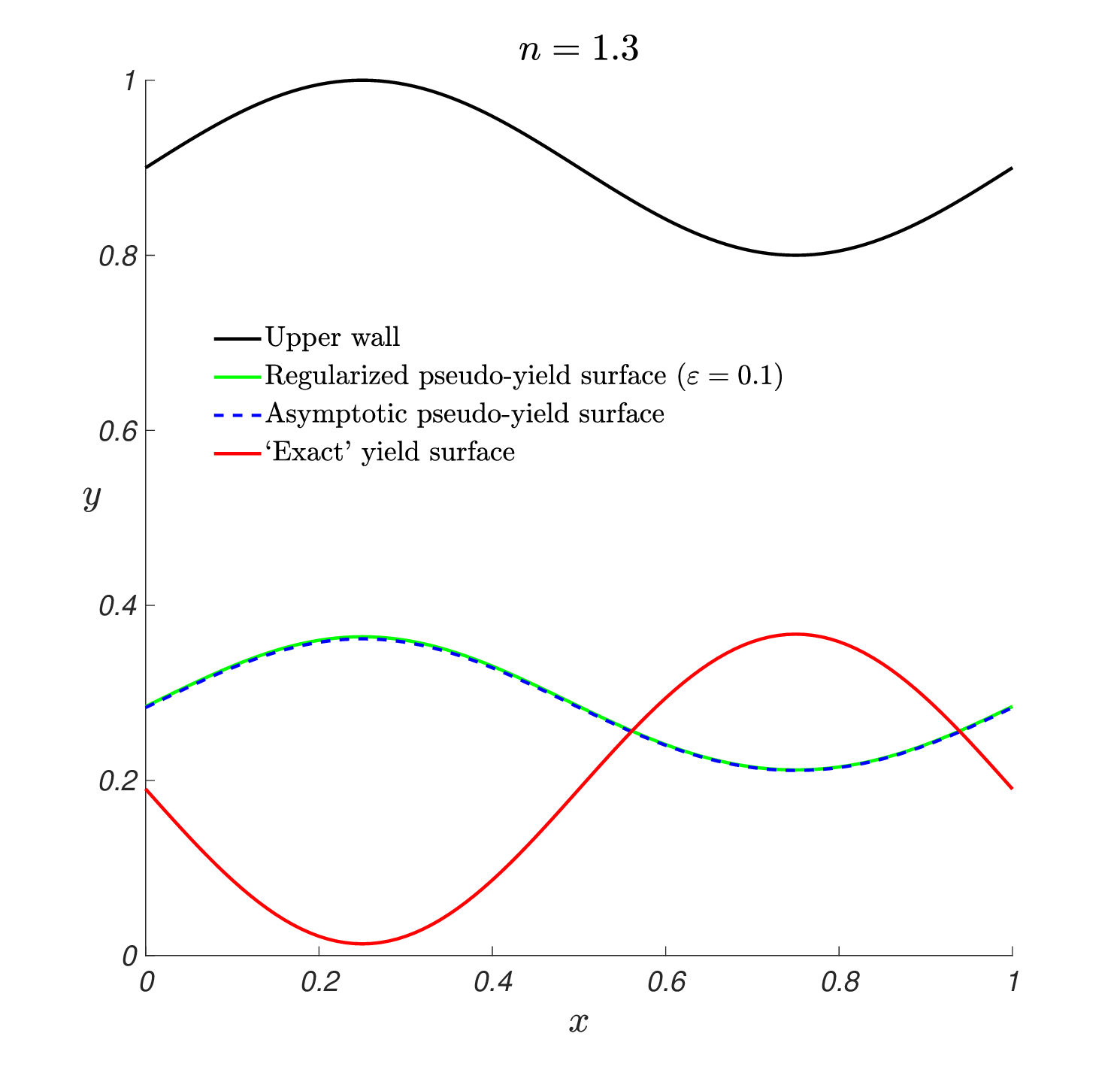}
         \caption{\label{pap13}}
         \end{subfigure}
        \begin{subfigure}{0.3\textwidth}
        \includegraphics[width=\textwidth]{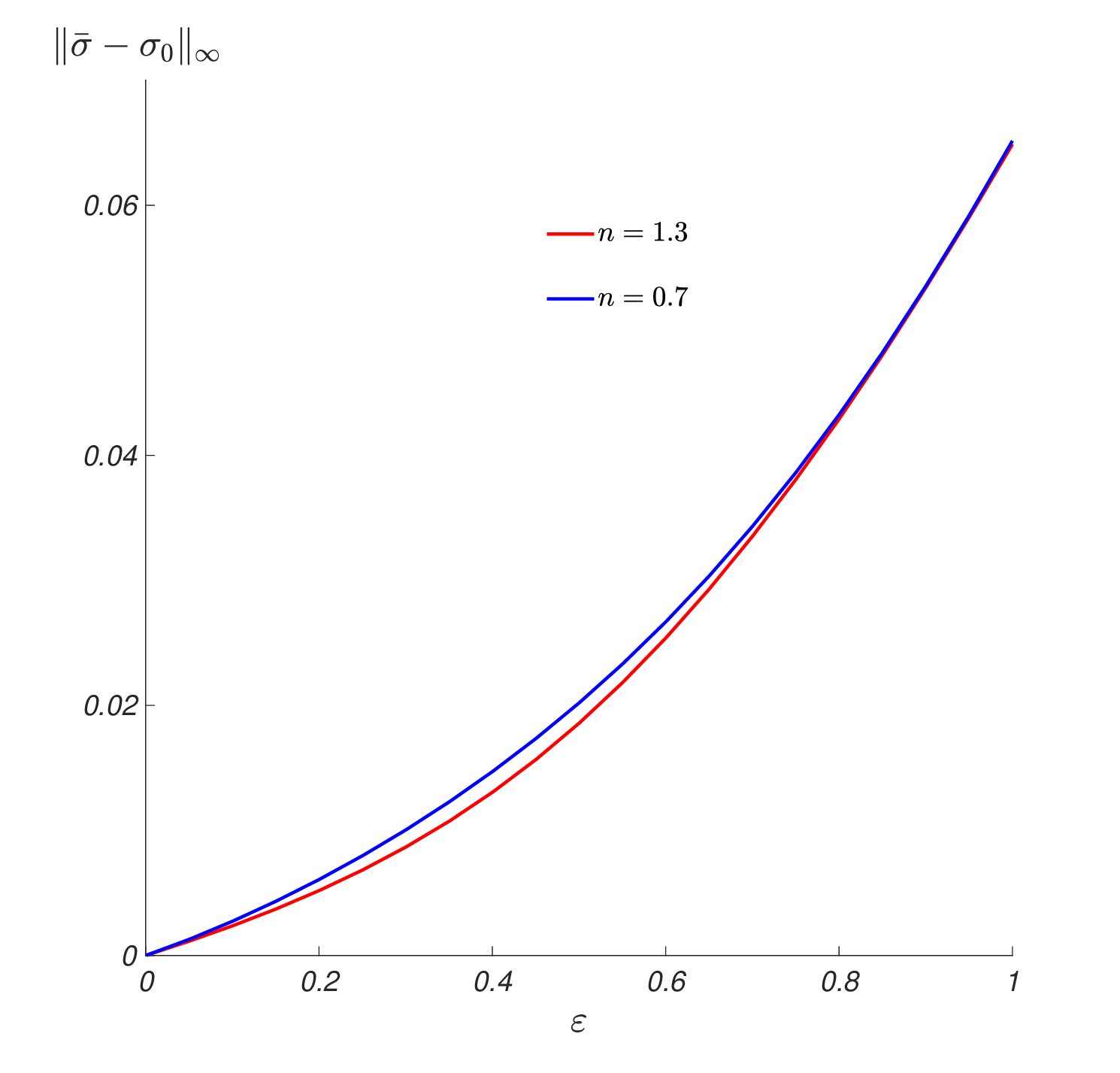}
        \caption{\label{papd}}
        \end{subfigure}
    \caption{Regularized and asymptotic pseudo-yield surfaces and `exact' yield surface at two distinct values of the power-law index $n$.  Here, we have considered a channel whose upper boundary has equation $h = 0.9+0.1\sin (2\pi x)$ and the Bingham number is equal to 3. The regularized pseudo-yield surface has been obtained by using the modified  Herschel-Bulkley-Papanastasiou model \eqref{papa}. To quantify somehow the rate of convergence of the regularized pseudo-yield surface to the asymptotic one, Figure \ref{papd} displays the sup-norms of the differences between $\bar{\sigma}$ and $\sigma_0$ as  functions of the regularization parameter $\veps$ for the cases analysed in \ref{pap07} and \ref{pap13}.}
    \label{papareg}
\end{figure}

\begin{figure}[h!]
    \centering
    \begin{subfigure}{0.3\textwidth}
         \includegraphics[width=\textwidth]{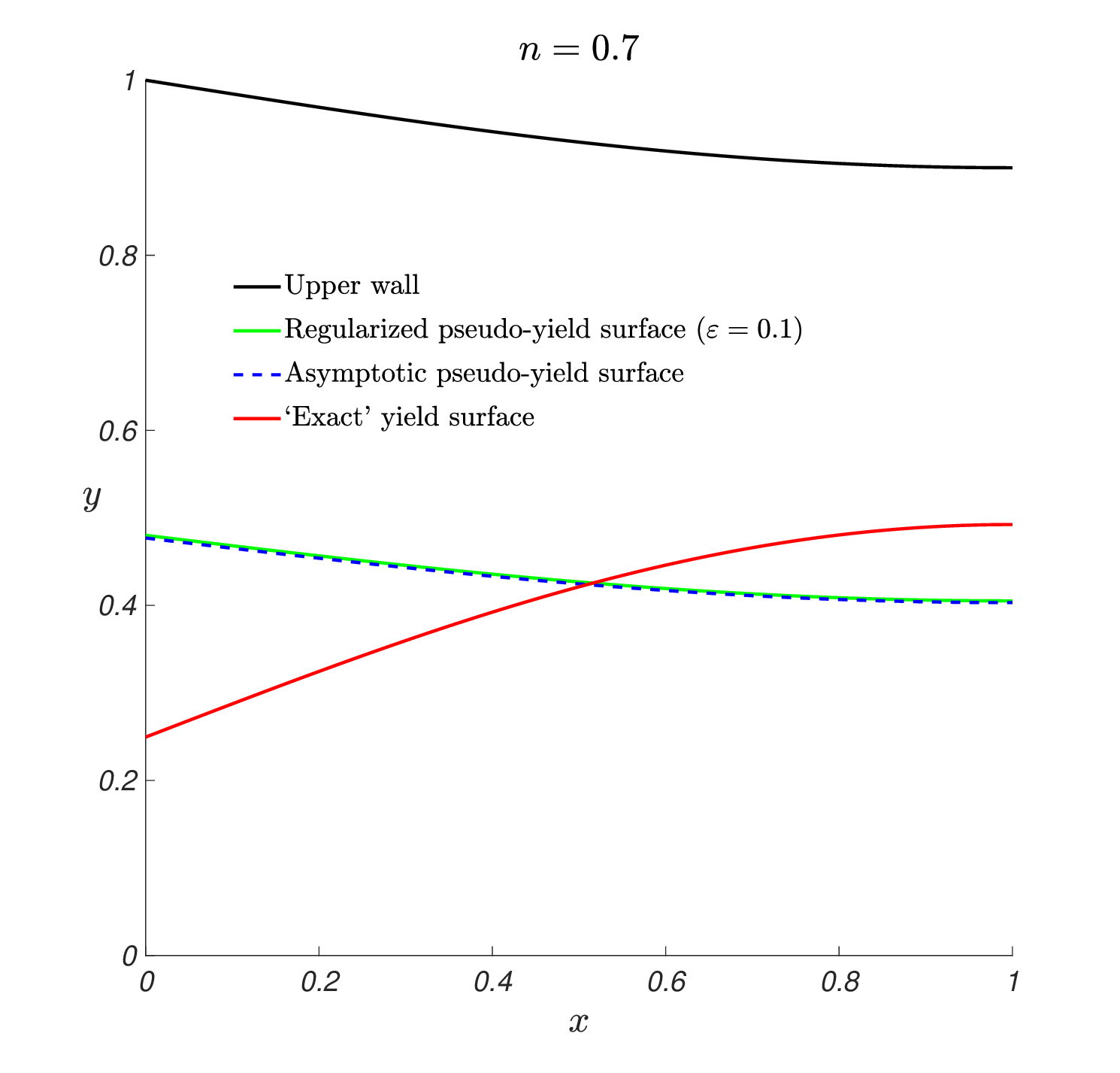}
         \caption{\label{ber07}}
         \end{subfigure}
         \begin{subfigure}{0.3\textwidth}
         \includegraphics[width=\textwidth]{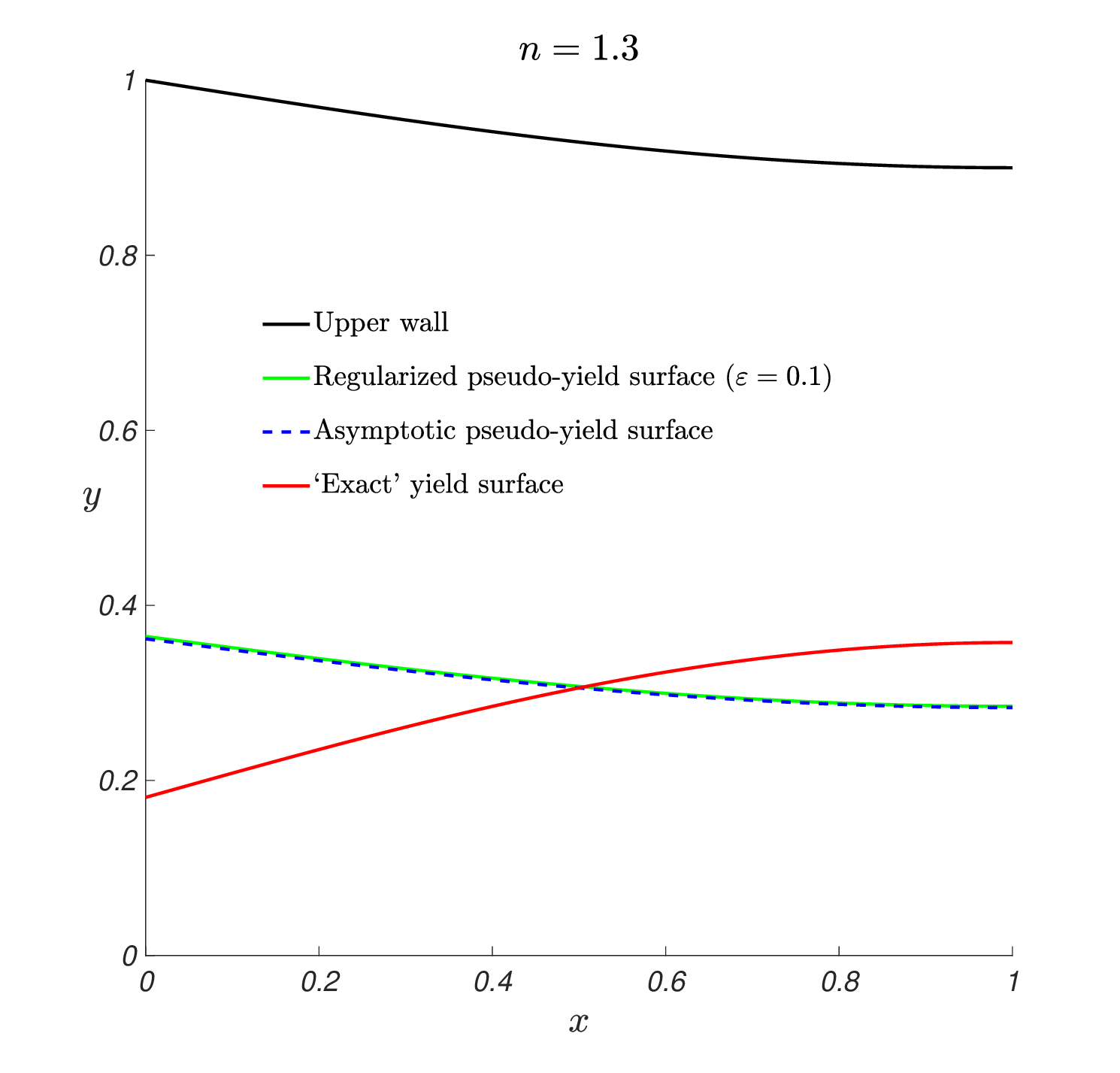}
         \caption{\label{ber13}}
         \end{subfigure}
        \begin{subfigure}{0.3\textwidth}
        \includegraphics[width=\textwidth]{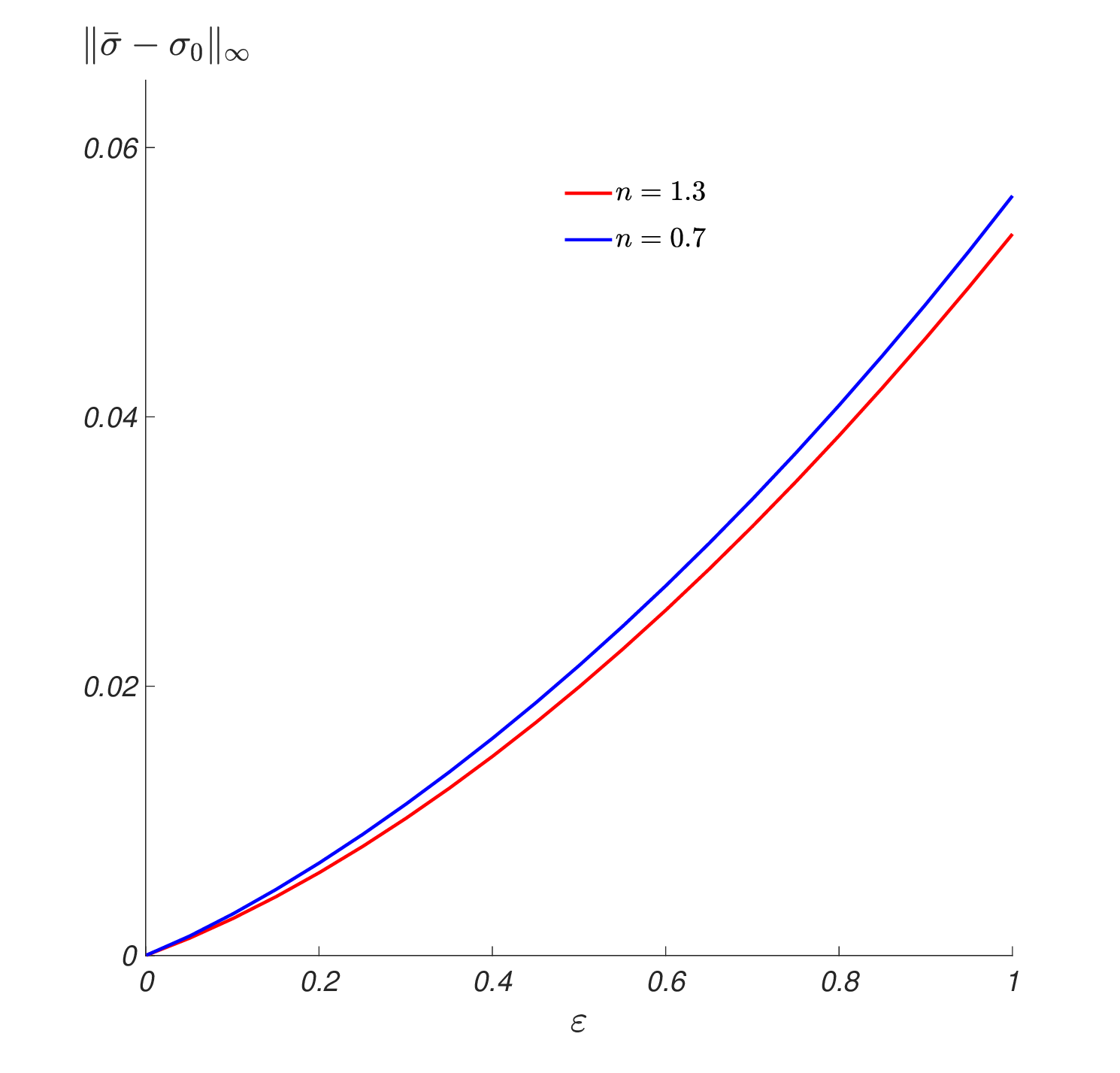}
        \caption{\label{berd}}
        \end{subfigure}
    \caption{Regularized and asymptotic pseudo-yield surfaces and `exact' yield surface at two distinct values of the power-law index $n$.  In this figure, we have considered a channel whose upper boundary has equation $h = 1+0.1\cos((\pi /2)(x+1))$ and the Bingham number is equal to 3. The regularized pseudo-yield surface has been obtained by using the modified Bercovier-Engelman regularization  \eqref{beren}. To quantify somehow the rate of convergence of the regularized pseudo-yield surface to the asymptotic one, Figure \ref{berd} displays the sup-norms of the differences between $\bar{\sigma}$ and $\sigma_0$ as  functions of the regularization parameter $\veps$ for the cases analysed in \ref{ber07} and \ref{ber13}.}
    \label{berreg}
\end{figure}

We have thus proven that, in the limit as the regularization parameter tends to zero, the yield surface and the components of the two-dimensional velocity field converge uniformly to continuously differentiable fields whose expressions depend exclusively on the profile of the boundaries of the channel and do not keep track of the particular regularized model used. 
These asymptotic results are then \emph{universal} for the class of regularized Herschel-Bulkley models \eqref{hbgen}. 
Moreover, for small values of the dimensionless regularization parameter $\veps$, the regularized pseudo-yield surface is practically identical to the asymptotic  surface $y=\sigma_0(x)$ (see Figures \ref{pap07}, \ref{pap13}, \ref{ber07} and \ref{ber13}). In Figures \ref{papd} and \ref{berd} we plot the sup-norms of the difference between the regularized pseudo-yield and asymptotic surfaces,
\be
\|\bar{\sigma}(\cdot;\veps)-\sigma_0\|_\infty=\max_{x\in[0,1]}|\bar{\sigma}(x;\veps)-\sigma_0(x)|,
\en
versus the dimensionless regularization parameter $\veps$ to give an idea of the rate of convergence of the regularized pseudo-yield surface to the asymptotic one. 

\begin{figure}[h!]
    \centering
         \begin{subfigure}{0.45\textwidth}
         \includegraphics[width=\textwidth]{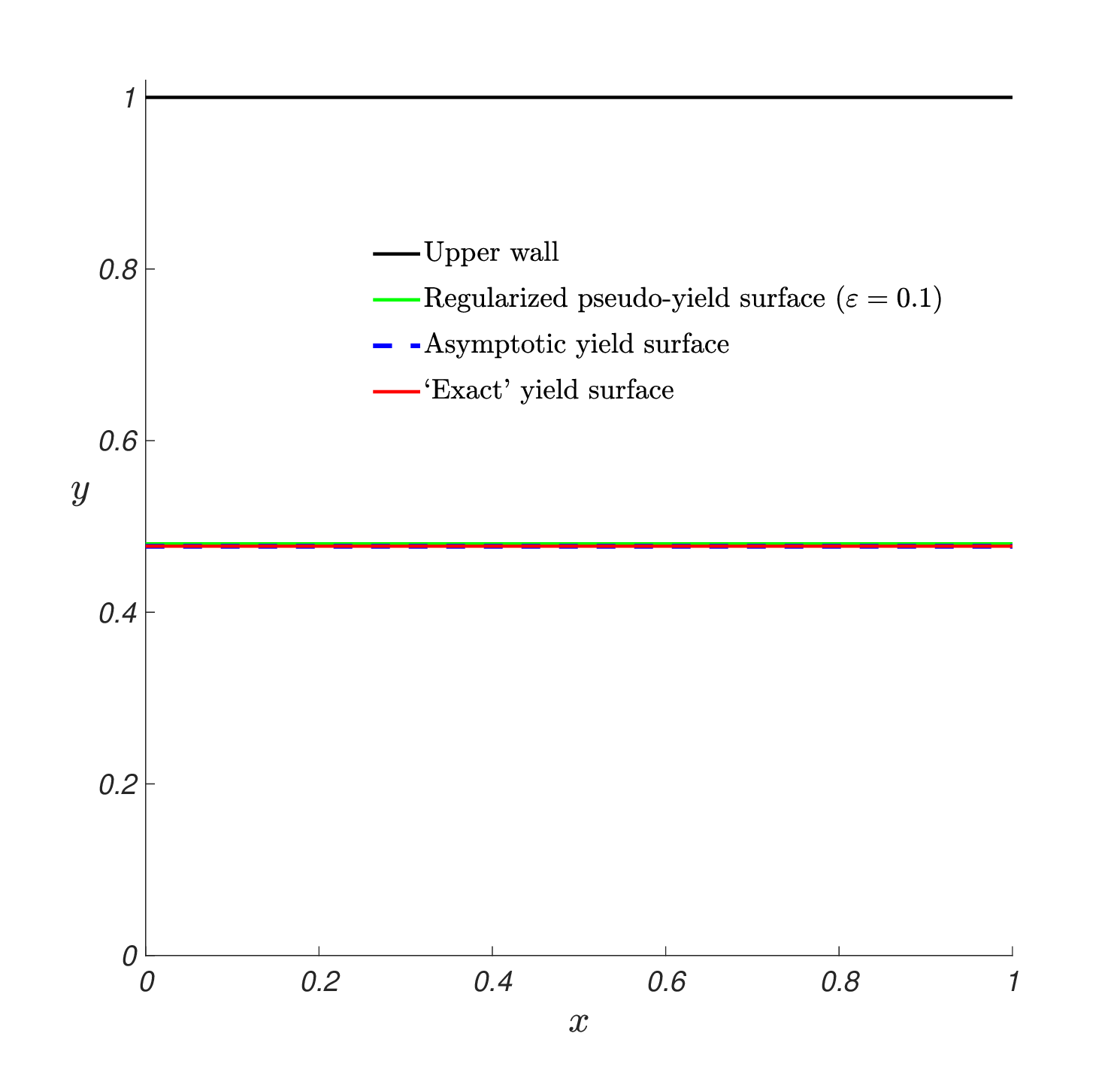}
         \caption{Model \eqref{papa} with $n=0.7$ \label{papapl}}
         \end{subfigure}
         \hfill
         \begin{subfigure}{0.45\textwidth}
         \includegraphics[width=\textwidth]{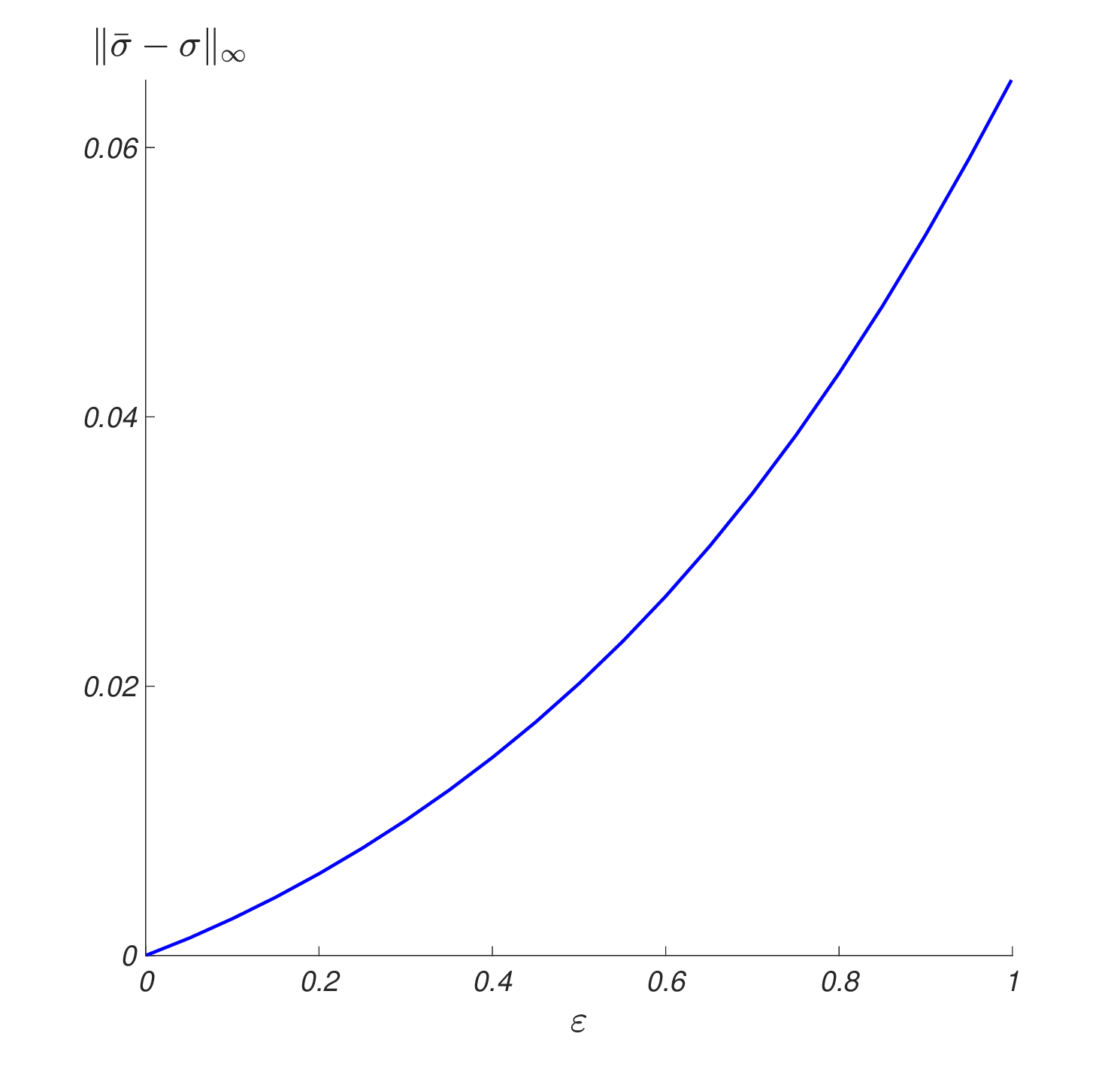}
         \caption{Model \eqref{papa} with $n=0.7$ \label{papapld}}
         \end{subfigure}
         \begin{subfigure}{0.45\textwidth}
         \includegraphics[width=\textwidth]{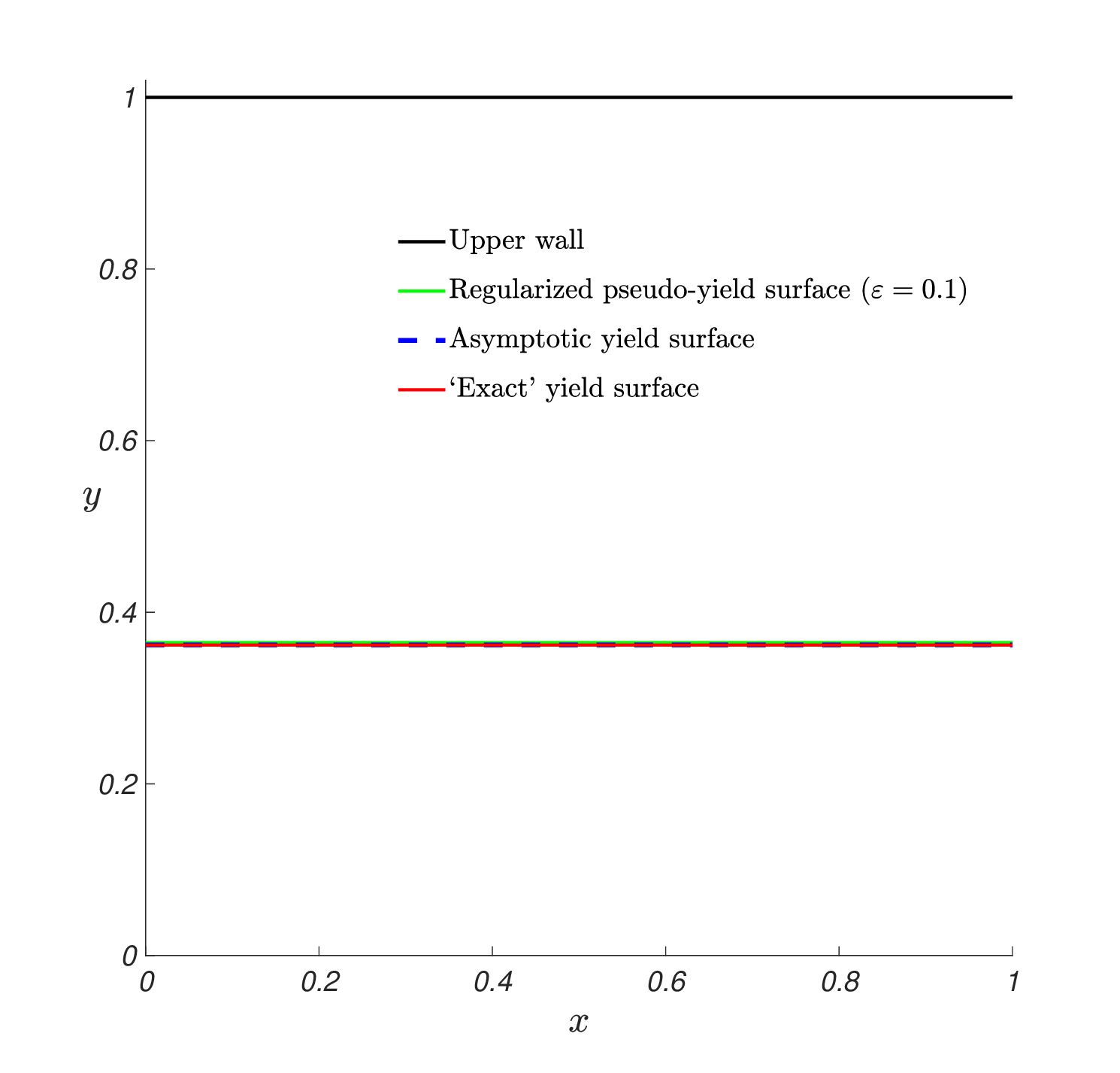}
         \caption{Model \eqref{beren} with $n=1.3$\label{berpl}}
         \end{subfigure}
         \hfill
         \begin{subfigure}{0.45\textwidth}
         \includegraphics[width=\textwidth]{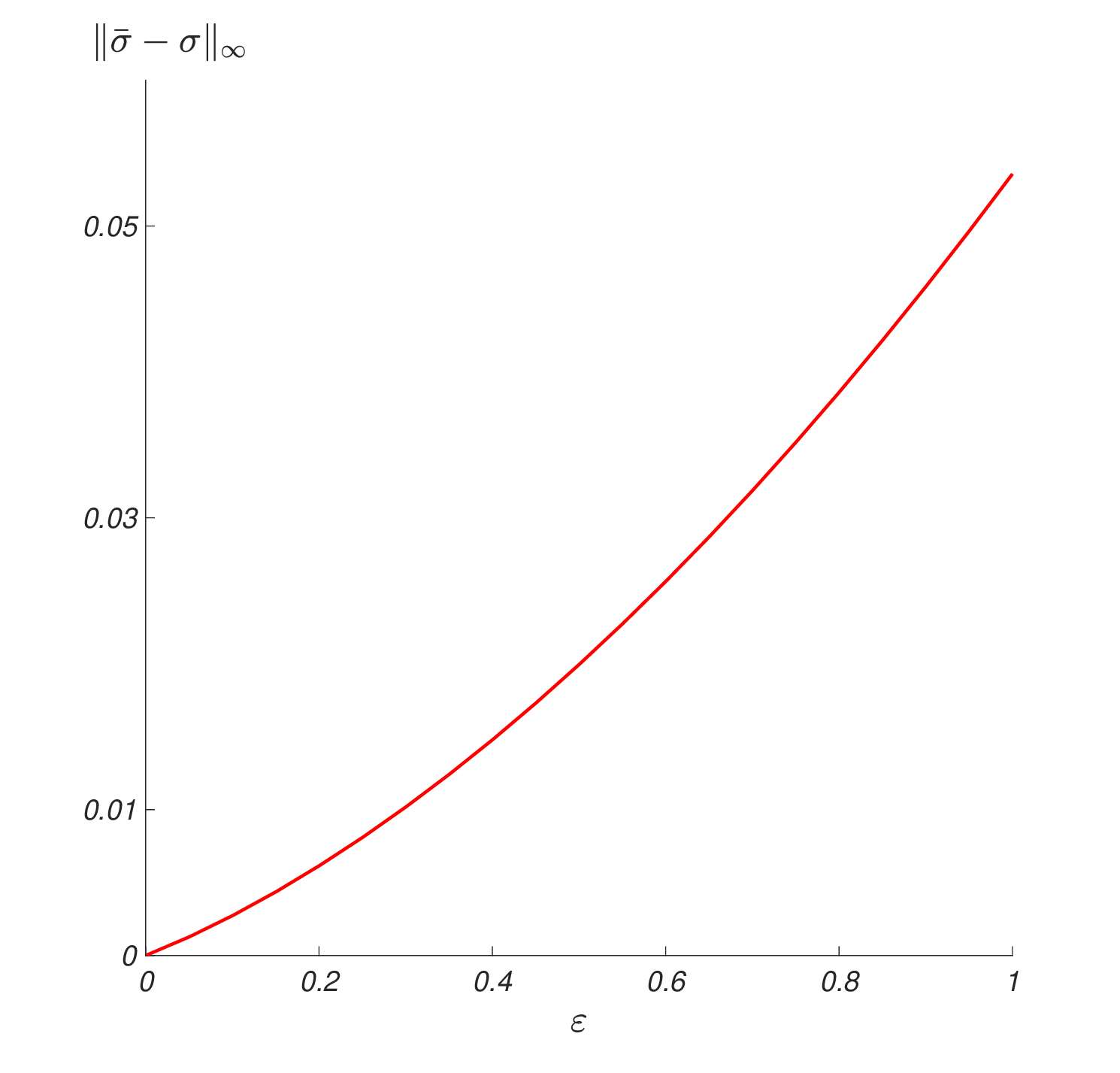}
         \caption{Model \eqref{beren} with $n=1.3$ \label{berdld}}
         \end{subfigure}
    \caption{Regularized pseudo-yield plane, `exact' and asymptotic yield planes in a plane channel at $Bm=0.3$.  The regularized pseudo-yield planes have been obtained by using the regularization \eqref{papa}, with $n=0.7$, in \ref{papapl}, and the regularization \eqref{beren}, with $n=1.3$, in \ref{berpl}. In both cases the regularization parameter $\veps$ has been taken equal to $0.1$. To quantify somehow the rate of convergence of the regularized pseudo-yield plane to the `exact' one, Figures \ref{papapld} and \ref{berdld} display the sup-norms of the differences between $\bar{\sigma}$ and $\sigma$ as  functions of the regularization parameter $\veps$ for the cases analysed in \ref{papapl} and \ref{berdld}, respectively.
    \label{papaberplane}}
\end{figure}

Observe that for non-plane channels the asymptotic surface $y=\sigma_0(x)$ does not coincide with the yield surface \eqref{sighb} predicted by the `exact' Herschel-Bulkley model. Indeed, resuming the discussion in the previous section, the asymptotic surface has the same monotonicity as the profile of the upper wall of the channel, whereas the `exact' yield surface has the opposite monotonicity of the boundary $y=h(x)$. Only for plane channels (for which $h\equiv1$) the asymptotic surface and the asymptotic two-dimensional flow coincide with the `exact' yield surface and `exact' velocity field (see Figure \ref{papaberplane}). To prove rigorously such a claim, it suffices to note that if $h\equiv1$, then the unknown $\vartheta$ in \eqref{eqay} is no more a function of the spatial variable $x$ and with the change of variables 
\be\label{eqs1}
\vartheta=\dfrac{2n+1}{n u_c}-\dfrac{n+1}{n} ,
\en
equation \eqref{eqay} becomes
\be
\dfrac{(n+1)^nu_c^{2n}[2n+1-(n+1)u_c]}{(2n+1)^{n+1}(nu_c-1)^{n+1}}=Bm ,
\en
that coincides with the equation for the plug-speed in plane channels, i.e. equation \eqref{uceq}, with $h\equiv1$. This implies that, for plane channels, the equation of the asymptotic yield plane can be expressed in terms of the plug-speed (cf. \eqref{eqs1}), whence inserting this expression in \eqref{u0} and \eqref{v0} gives, respectively, the components of the `exact' velocity field \eqref{uf} and \eqref{vf} when $h\equiv 1$.

\begin{figure}
\centering
\begin{subfigure}{0.45\textwidth}
    \includegraphics[width=\textwidth]{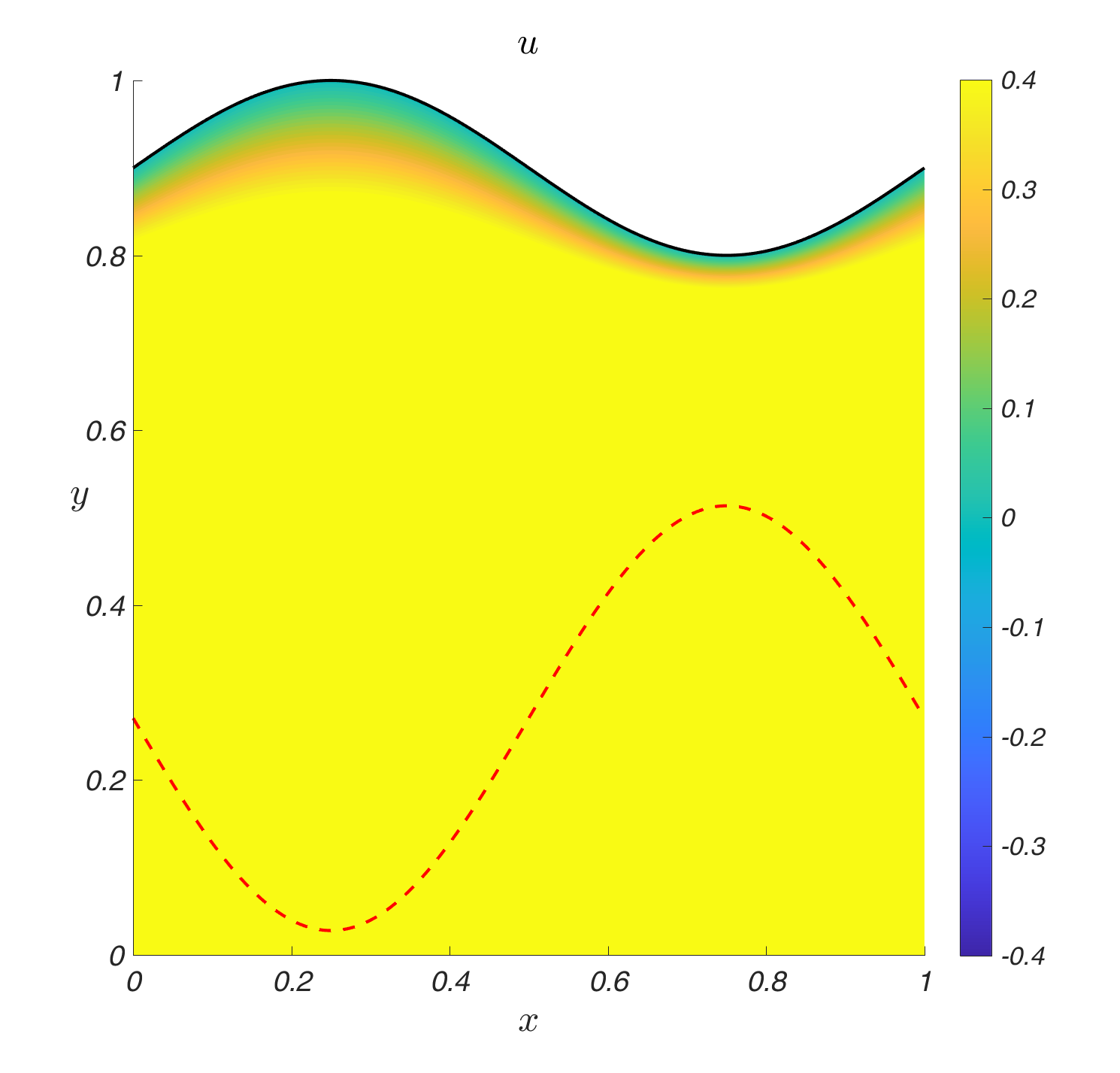}
    \caption{\label{ues}}
\end{subfigure}
\hfill
\begin{subfigure}{0.45\textwidth}
    \includegraphics[width=\textwidth]{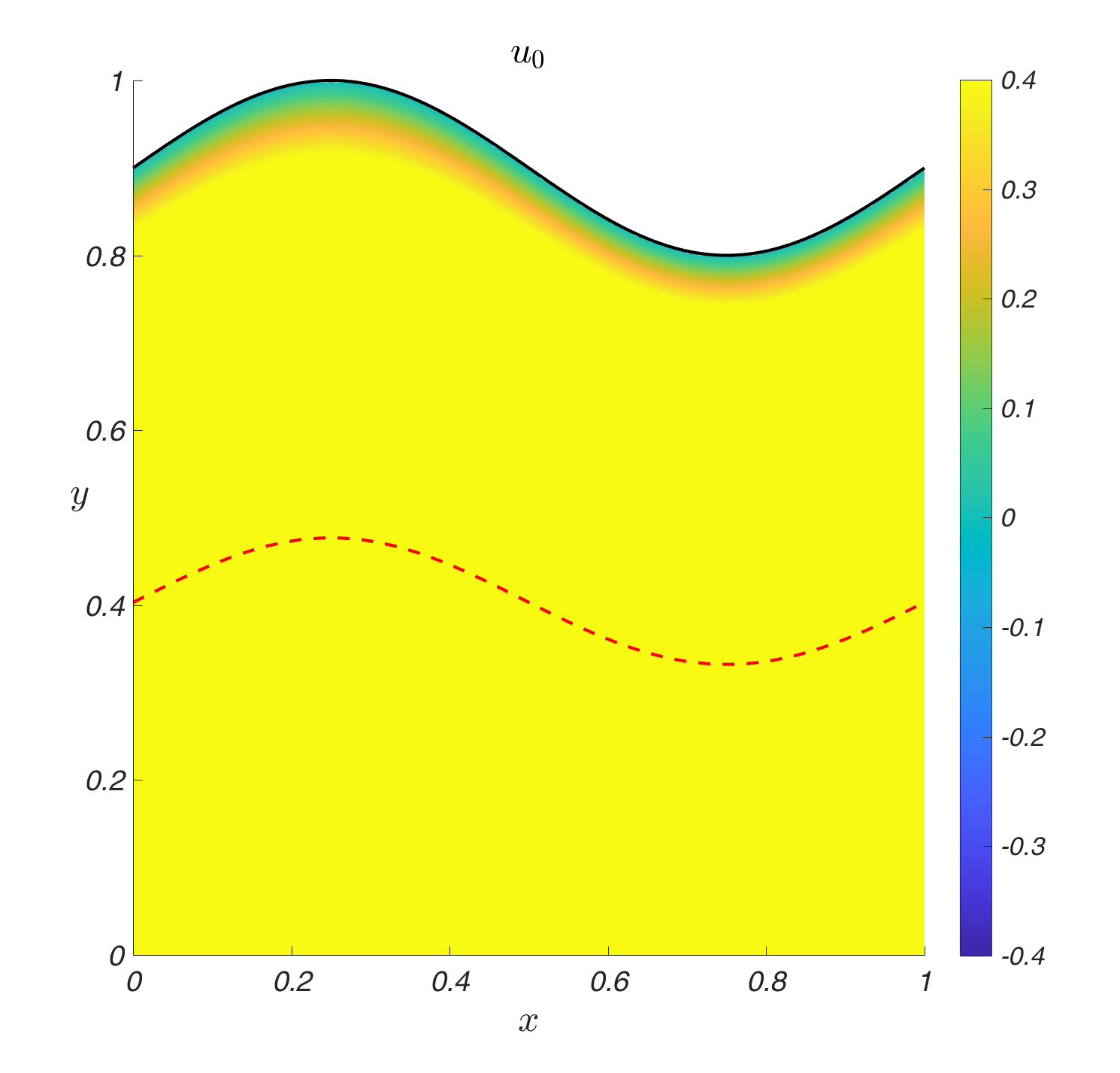}
    \caption{\label{uasy}}
   \end{subfigure}
    \begin{subfigure}{0.45\textwidth}
    \includegraphics[width=\textwidth]{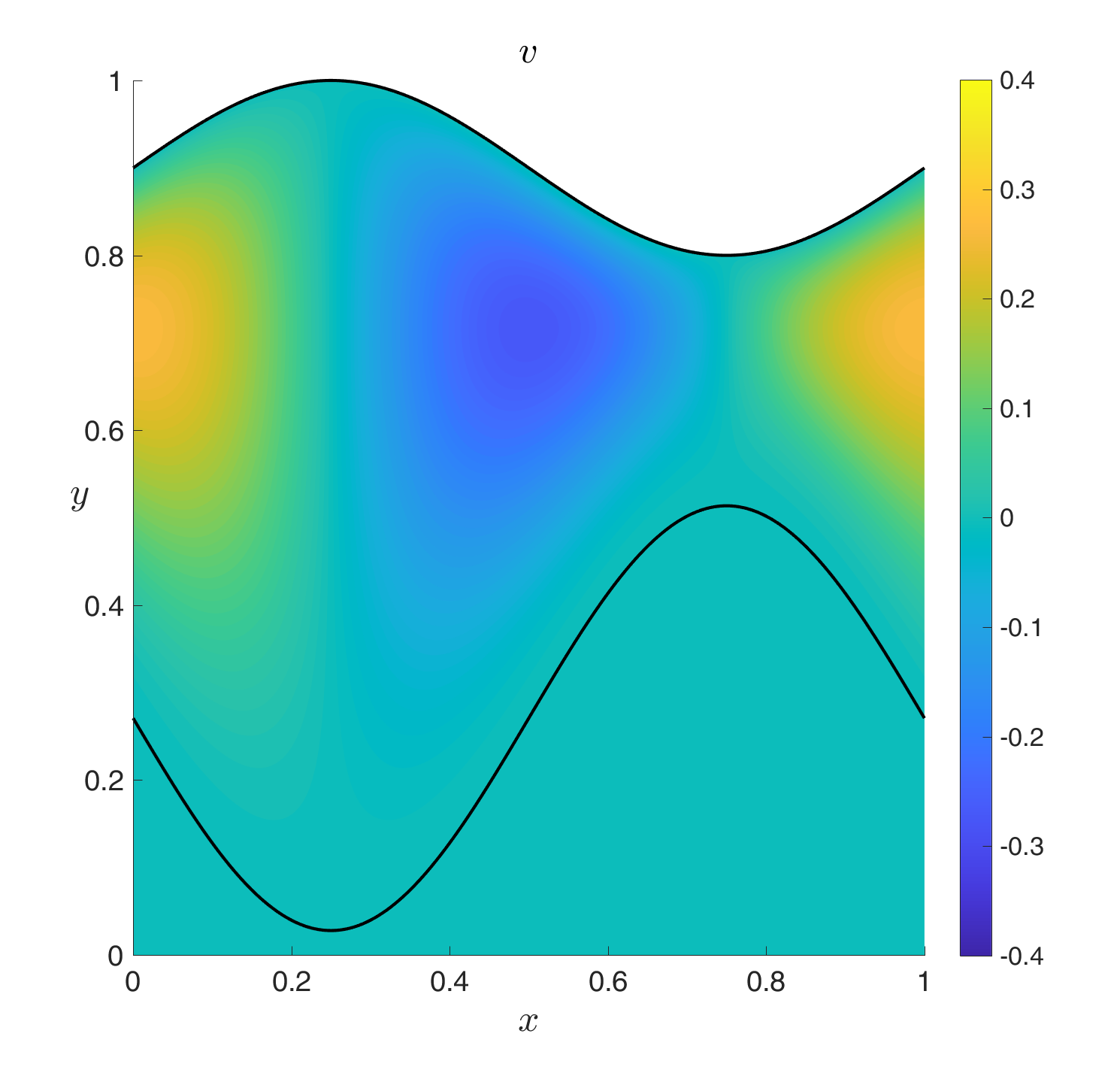}
    \caption{\label{ves}}
\end{subfigure}    
\hfill
 \begin{subfigure}{0.45\textwidth}
    \includegraphics[width=\textwidth]{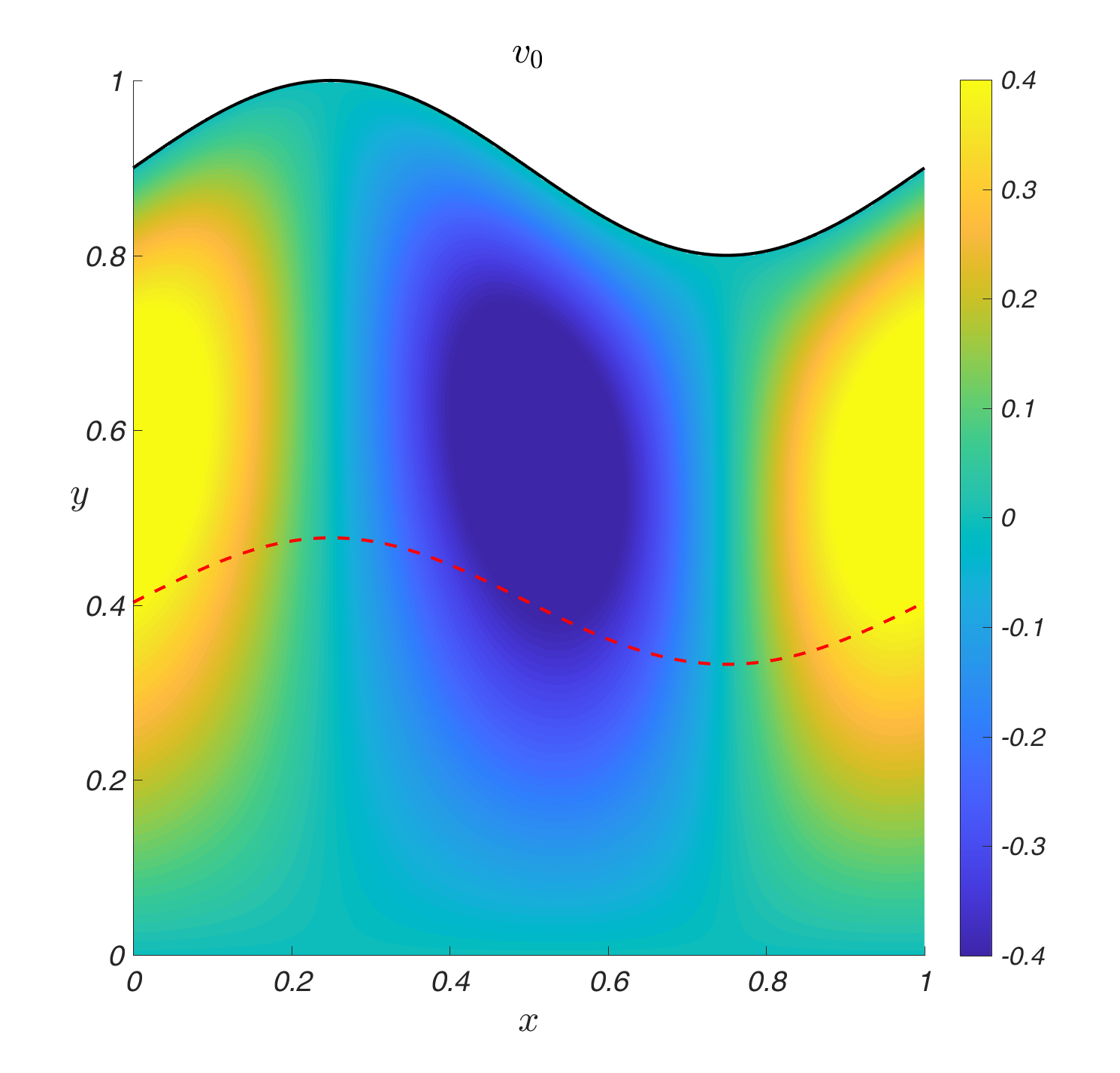}
    \caption{\label{vasy}}
\end{subfigure}  
\caption{Comparison between the components of the velocity field predicted by the Herschel-Bulkley model (\ref{ues} and \ref{ves}) vs those determined  by  any regularized model in the limit as $\veps\rightarrow0^+$ (\ref{uasy} and \ref{vasy}). Here, the profile of the upper wall (solid black lines) has equation $h(x)=0.9+0.1\sin(2\pi x)$, $Bm=3$ and $n=0.7$. The dashed red lines represent the `exact' yield surface  (\ref{ues} and \ref{ves}), or the asymptotic pseudo-yield surface (\ref{uasy} and \ref{vasy}). \label{velcomp}}
\end{figure}

Finally, note that, as in the`exact' unyielded region $\Omega_c$, from \eqref{su} and \eqref{lim1} the shear stress is indeterminate in 
\be
\Omega_0=\{(x,y)\in\Omega:\, |y|\leq \sigma_0(x)\},
\en
as the shear strain rate vanishes there. However, for non-plane channels, the motion of the fluid in $\Omega_0$ is not rigid as the velocity is not constant there (see Figure \ref{velcomp}). For non-plane channels, it is appropriate to regard the domain $\Omega_0$ as the \emph{asymptotic pseudo-yield region} and the loci $y=\pm\sigma_0(x)$ as the \emph{asymptotic pseudo-yield surfaces}. Only for plane channels for which the asymptotic and `exact' yield surfaces and velocity fields coincide, the motion of the fluid in $\Omega_0$ is rigid (c.f. \eqref{u0} and \eqref{v0} with $h\equiv1$) and $\Omega_0=\Omega_c$. This means that the flow of an `exact' Herschel-Bulkley fluid can be retrieved from any of its regularizations in the limit as the regularization parameter tends to zero only if the channel is plane.


\section{Concluding remarks\label{Sec:Dis+Conc}}
We have studied the steady two-dimensional flows of an incompressible viscoplastic fluid in a symmetric channel with impermeable and no-slip boundaries. The active part of the stress tensor has been considered to be given by the Herschel-Bulkley model or by a regularization of its. In order to obtain flows in closed form, we have assumed that the aspect ratio of $\Omega$ is small enough to make sense to introduce the lubrication approximation. In so doing and following similar arguments as in \cite{Farina2024}, we have been able to solve the approximated boundary value problem that governs the symmetric steady 2D-flows in a viscoplastic fluid both when considering the `exact' Herschel-Bulkley model (with a non-smooth apparent viscosity) and when one introduces a regularized viscosity function. We have shown that, at small enough values of the regularization parameter, the regularized pseudo-yield and `exact' yield surfaces are almost identical only in plane channels. These two surfaces coincide in the limit as the regularization parameter tends to zero as a result of the coincidence between the asymptotic 2D flows \eqref{2de} and \eqref{2d0} when $h\equiv 1$. On the contrary, the flows predicted by the `exact' Herschel-Bulkley model and its regularizations are markedly different in channels with curved walls, also in the limit as $\veps\rightarrow0^+$ (Figure \ref{velcomp}). Consequently, for non-plane channels the `exact' unyielded region and the asymptotic pseudo-unyielded region differ not only quantitatively but also qualitatively. Indeed, the `exact' yield surfaces have the opposite monotonicity of the profiles of the walls  of the channel, while the asymptotic pseudo-yield surfaces have the same monotonicity as the boundaries $y=\pm h(x)$.

Finally, our asymptotic analysis has revealed that, irrespective of the functional form of the regularization of the Herschel-Bulkley models considered, in the limit as the regularization parameter tends to zero the asymptotic 2D flows (and thus the pseudo-yield surfaces) depend exclusively on the profiles of the boundaries of the channel and the value of the Bingham number. These results are then universal for the class of regularizations \eqref{hbgen}-\eqref{regcond}. Farina \emph{et al} \cite{Farina2024} obtained similar results for Bingham fluids, i.e. when considering the model \eqref{hbmod} with $n=1$. We can then conclude that the universal results obtained here generalize those for Bingham fluids to Herschel-Bulkley fluids.


\section*{Acknowledgments}
B.C., A.F. and L.F. have been funded by the NextGenerationEU PRIN 2022 research project ‘Mathematical Modelling of Heterogeneous Systems’ (grant n. 2022MKB7MM). L.V. acknowledges the funding from Project ‘Mathematical modelling for a sustainable circular economy in ecosystems’ (grant n. P2022PSMT7) funded by EU in NextGenerationEU plan through the Italian ‘Bando Prin 2022 - D.D. 1409 del 14-09-2022’ by MUR. The Authors performed this study under the auspices of the GNFM of Italian INDAM.


\vskip2pc



\end{document}